\documentclass[10pt,twocolumn,article]{IEEEtran}

\usepackage{etex}
\usepackage{tikz}

\usetikzlibrary{shapes,arrows}

\usetikzlibrary{shapes,arrows,trees,calc}

\usetikzlibrary{arrows,shapes,positioning,shadows,trees}

\tikzset{
  basic/.style  = {draw, text width=2cm, drop shadow, font=\sffamily, rectangle,},
  root/.style   = {basic, rounded corners=2pt, thin, align=center,
                   fill=green!30},
  level 2/.style = {basic, rounded corners=6pt, thin,align=center, fill=red!20,
                   text width=12em},
  level 3/.style = {basic, thin, align=left, fill=olive!15, text width=8.5em}
}
\usepackage[T1]{fontenc}
\usepackage{tikz-qtree}
\usepackage{epigraph}
\usepackage{adjustbox}
\usepackage{setspace}
\usepackage{cite}
\usepackage{epsf}\epsfverbosetrue
\usepackage{graphics,epsfig,color}
\usepackage{graphicx,epsfig}
\usepackage{epsfig}
\usepackage{multirow}
\usepackage{slashbox}
\usepackage{alltt}
\usepackage{subfigure}
\usepackage{color}
 \usepackage{float} 
\usepackage{url}
\usepackage{graphicx}
\usepackage{verbatim} 
\usepackage{footnote} 
\usepackage{latexsym} 
\usepackage{amsmath}
\usepackage{sidecap} 
\usepackage{color}
\usepackage{wrapfig}
\usepackage{algorithm}
\usepackage{eso-pic}
\usepackage{fix-cm}
\usepackage{mcite}

\usepackage{color}
\usepackage{wrapfig}
\usepackage{algorithm}
\usepackage{layout}
\usepackage{amsmath}
\usepackage{textcomp}
\usepackage{multirow}
\usepackage{soul}
\usepackage{amsmath}
\usepackage{verbatim}
\usepackage{smartdiagram}
\usepackage{metalogo}
\usepackage{bbding}  
\usepackage{amsfonts,amssymb}

\usepackage{tikz}
\usetikzlibrary{shapes}
\usepackage{amsmath}
\usepackage{xspace}

\usepackage{tikz}
\makeatletter 
\@namedef{color@1}{red!40}
\@namedef{color@2}{green!40}   
\@namedef{color@3}{blue!40} 
\@namedef{color@4}{cyan!40}  
\@namedef{color@5}{magenta!40} 
\@namedef{color@6}{yellow!40}    

\newcommand{\graphitemize}[2]{%
\begin{tikzpicture}[every node/.style={align=center}]  
  \node[minimum size=5cm,circle,fill=gray!40,font=\Large,outer sep=1cm,inner sep=.5cm](ce){#1};  
\foreach \gritem [count=\xi] in {#2}
{\global\let\maxgritem\xi}  
\foreach \gritem [count=\xi] in {#2}
{%
\pgfmathtruncatemacro{\angle}{360/\maxgritem*\xi}
\edef\col{\@nameuse{color@\xi}}
\node[circle,
     ultra thick,
     draw=white,
     fill opacity=.5,
     fill=\col,        
     minimum size=3cm] at (ce.\angle) {\gritem };}%
\end{tikzpicture}  
}%

\newcommand{\blue}[1]{\textcolor{black}{#1}}

\newcommand{\amj}[1]{\textcolor{black}{#1}}

\newcommand*{\rom}[1]{\expandafter\@slowromancap\romannumeral #1@}
\newcommand{\Rmnum}[1]{\expandafter\@slowromancap\romannumeral #1@}
\makeatother

\usepackage[utf8]{inputenc}
 \usepackage{dtklogos}
 \usepackage{tikz}
 \usetikzlibrary{mindmap,shadows}
\usepackage[hidelinks,pdfencoding=auto]{hyperref}

\begin{document}
\setstcolor{red}
\title{Effective Capacity in Wireless Networks: A Comprehensive Survey}

\author{Muhammad Amjad, Leila Musavian, and Mubashir~Husain~Rehmani
\thanks{Please direct correspondence to M. Amjad.}
\thanks{M. Amjad and L. Musavian are with the School of Computer Science and Electronic Engineering, University of Essex, CO4 3SQ, UK (Email: m.amjad@essex.ac.uk, leila.musavian@essex.ac.uk)}
\thanks{M.H.~Rehmani is with Department of Computer Science, Cork Institute of Technology, Rossa Avenue, Bishopstown, Cork, Ireland. (Email:mshrehmani@gmail.com).}}

\maketitle

\begin{abstract}
\amj{Low latency applications, such as  multimedia communications, autonomous vehicles, and Tactile Internet are the emerging applications for next-generation wireless networks, such as 5th generation (5G) \blue{mobile networks}. Existing physical-layer channel models, however,  do not explicitly consider quality-of-service (QoS) aware related parameters  under specific delay constraints. To investigate the performance of low-latency applications in future networks, a new mathematical framework is needed. Effective capacity (EC), which is a link-layer channel model with QoS-awareness, can be used to investigate the performance of wireless networks under certain statistical delay constraints.}   In this paper, we provide a comprehensive survey on  existing works,  \amj{ that use the EC model} in various wireless networks. We summarize  \amj{the work related to} EC    for different networks such as cognitive radio networks (CRNs), cellular networks, relay networks, adhoc networks, and mesh networks.  We explore  five case studies encompassing  EC operation with different design and architectural requirements. We survey  various delay-sensitive applications such as voice and video  with their EC analysis under certain delay constraints. We  finally present the future research directions with open issues covering  EC maximization.      
\end{abstract}


\section{Introduction}
\label{sec:introduction}

\subsection{Motivation: Need of Effective Capacity Mathematical Model in Wireless Communications}
\label{sec:motiv}
Advances in wireless communications have resulted into emergence of a wide range of applications.  Emerging wireless networks with advanced technologies such as full-duplex (FD) communications, non-orthogonal multiple access (NOMA), multiple input and multiple output (MIMO) and millimeter wave (mmWave)  promise  higher data rates \cite{R71,R230,R234}. With  provision of this higher data rate and seamless connectivity, multimedia applications, \blue{which  are regarded as  delay-sensitive applications,} have gained a lot of attraction  \cite{amjad1}. This requires \amj{ an}  efficient modeling of wireless channel that can take into consideration  QoS metrics such as delay-violation probability, data rate, and end-to-end delay \cite{R132}. 

Packet switched networks can be analysed with the help of physical and link-layer channel models \blue{depicted}  in  Figure \ref{fig:ec-packet-switched-networks}.   \amj{Using}  physical-layer channel models   for analysing the performance of delay-limited  \blue{applications}   \amj{can be} complex and inaccurate \amj{in some cases}    \cite{R110}. Hence,  a new link-layer channel model named as ``effective capacity (EC)'' has been introduced  \cite{R110}. With the help of EC,  the channel can be modeled in terms of \amj{link-layer related} QoS-metrics, such as   \amj{probability of having non-empty buffer} \blue{and} delay violation probability.  Concept of this link-layer channel model was first introduced in \cite{R110},  \amj{which}  modeled \amj{a wireless link}  using two EC functions named as QoS exponent  and probability of non-empty buffer.  \amj{The developed}    link-layer channel model  provides advantages such as  ease   of implementation and translation  into the QoS guarantee, i.e., delay violation probability.  Main motivations involving  EC metric for various performance evaluations  \blue{are}   highlighted below:

\begin{figure*}  
\centering
  \includegraphics[width=5cm,height=6.5cm]{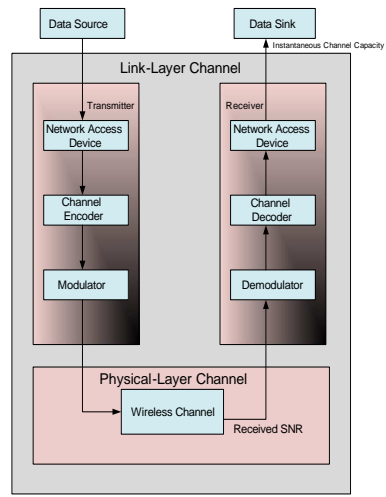}
\caption{ Basic components involved in the communications of packet switched networks \protect\cite{R110}. In this packet-based communications system,  different components of physical and link-layer have been illustrated which shows the difference between  physical and link-layer channel modelling.    }
\label{fig:ec-packet-switched-networks}
\end{figure*}
\begin{itemize}
 
\item EC modelling is based on  an in-depth queueing analysis which can be used to   characterize \amj{a relation between}  \blue{the} source \amj{rate  and \blue{the} service rate taking into consideration both link-layer and physical layer parameters}. Through this  characterization, advance validation of communications systems \blue{performance} such as  \amj{efficient} admission control  can  be achieved \cite{R49}.

\item EC is  the  dual  \amj{concept} of effective bandwidth \cite{R81,R305} and  \amj{shows}  the maximum constant arrival rate for a wireless channel while satisfying  \amj{a delay outage probability constraint} \cite{R22}. This feature can be exploited to achieve the required QoS for some applications with specific QoS requirements.

\item With the help of the EC concept,  QoS provisioning over wireless links and  \amj{efficient}  bandwidth allocation can be achieved in \amj{closed-form} while satisfying  certain delay guarantee constraints \cite{R110}.

 \item  \amj{The EC performance of well-known}  physical layer-\amj{based resource allocation algorithms, e.g, water filling,}  can be investigated.   Performance of various proposed adaptive modulation and coding (AMC)  \amj{schemes}\footnote{For further  details,  see Sections \ref{sec:path-loss-models} and \ref{sec:retransmission}.}  can  be  \amj{tested}  by using \blue{the} EC metric \cite{R08}. 
 
 \item   \blue{Using the EC model, the performance of}   \amj{adaptive resource}   allocation \blue{techniques}   \amj{for} a specific QoS-aware connection can  be analysed  \amj{in closed-form} \blue{in various cases}. \amj{This, in turn, will pave the way for \blue{designing efficient resource allocation algorithms, hence improving} }    \amj{ the system performance.}

 \item Provision of QoS guarantee with support for a variety of traffic flows requires  efficient scheduling techniques. Using the EC concept, efficient delay constrained scheduling approaches can be designed  \cite{R66}.
 
\end{itemize}

\subsection{Contribution of This Survey Article}
The concept of EC has been extensively used in  literature to test the performance of various delay-constrained wireless networks.
However, to the best of our knowledge, there exist no comprehensive survey that can encompass  the state-of-the-art work of the EC model. This survey paper will be  \amj{a}  first attempt to provide  \amj{a} comprehensive view of  the EC  \amj{model}  in wireless communications systems. In summary, following are the core contributions of our work: 

\begin{itemize}
 
 \item We cover a broad description of various applications  \amj{for which their performance}   can be analysed using \blue{the  EC model}. 
 \item We discuss  five  case studies that highlight the use of  EC in five different wireless networks. 
 \item We survey the achievable EC  \amj{for various types of fading Models}. 
 \item We survey the achievable EC \blue{of cognitive radio networks, wireless local area networks, mesh networks, cellular networks, and    full-duplex (FD) communications.} 
 \item We outline  future research directions and open issues related to our survey, i.e., how the concept of EC can be used to analyze the performance   \amj{of various}  wireless networks, \amj{and how this concept can be used for proposing efficient resource allocation and transmission designs.} 
\end{itemize}

\begin{table}
\footnotesize
\centering
\caption{List of acronyms and corresponding definitions.}
\label{acronym}
\begin{tabular}{|p{2cm}|p{4.5cm}|}
\hline
\bfseries Acronyms & \bfseries Definitions \\
\hline
   5G &	Fifth Generation\\\hline
  AF & Amplify-and-Forward        \\\hline
  AMC &  Adaptive Modulation and Coding        \\\hline
  ARQ &  Automatic Repeat reQuest        \\\hline
   AWGN& Additive White Gaussian Noise         \\\hline
  BER & Bit Error Rate         \\\hline 
   BS&  Base Station        \\\hline
  CDMA &Code-Division Multiple Access          \\\hline
   CR&  Cognitive Radio         \\\hline
   CRN& Cognitive Radio Networks         \\\hline
   CSI&   Channel State Information       \\\hline 
   DF&   Decode-and-Forward       \\\hline
   DSA&   Dynamic Spectrum Access      \\\hline
   EC&   Effective Capacity       \\\hline
   FIFO&  Ftrst-In-First-Out        \\\hline
 FDCRNs &   Full-Duplex Cognitive Radio Networks       \\\hline   
   HARQ&     Hybrid  Automatic Repeat reQuest     \\\hline
   IP&  Internet Protocol        \\\hline 
   LLC&   Logical Link Control        \\\hline
   LTE&     Long-Term Evolution      \\\hline
   MAC&  Media Access Control address        \\\hline
   MIMO&Multiple-Input and Multiple-Output          \\\hline
   MS&   Mobile Stations       \\\hline 
   NC&    Network Coding      \\\hline
  OFDMA & Orthogonal Frequency-Division Multiple Access       \\\hline
   PU&  Primary User        \\\hline
   PLR&   Packet Loss Ratio       \\\hline
   QoE& Quality of Experience         \\\hline
   QoS& Quality of Service         \\\hline 
   SINR&   Signal-to-Interference-plus-Noise Ratio        \\\hline 
   SDR& Software-Defined Radio          \\\hline
  SUs &   Secondary Users       \\\hline 
  SNR &  Signal-to-Noise Ratio        \\\hline
   TCP&  Transmission Control Protocol       \\\hline
   TDMA&   Time-Division Multiple Access       \\\hline
   TVWS& TV White Spaces         \\\hline
   VANETs& Vehicular Ad hoc Networks         \\\hline
   VoD&  Video on Demand        \\\hline 
     WLAN                      & Wireless Local Area Networks   \\\hline
                      WRANs&Wireless Regional Networks    \\\hline
                      WSNs &Wireless Sensor Networks \\\hline
                                                 ZFBF&Zero-Forcing Beamforming \\\hline              
                                                           \end{tabular}
\end{table}

\subsection{Review of Related Survey Articles}
As far as we can determine, there exists no comprehensive survey, that  covers  EC studies in various wireless networks.  State-of-the-art  work using \blue{the  EC model} has been performed  with \amj{various}  fading conditions, supported applications, antenna designs, employed networks,  retransmission schemes, etc. Most of the work on EC only covers one or more aspects of communications. There exists very limited work that  covers  multiple aspects of EC regarding the provision of statistical QoS in wireless communications.  Authors in  \cite{R314}, have discussed various potentials and challenges that are associated with the   \blue{provisioning} of statistical QoS requirements in buffer-aided \amj{relay} communication systems. Using the concept of EC, a trade-off of statistical delay between two concatenated queues has been discussed.     \amj{This}  study  \amj{focuses on}  one and two \blue{hops} relay systems  \amj{and summarizes the performance in terms of achievable rate \blue{and}  secrecy rate.} \amj{The survey also covers the performance of caching in delay QoS-aware relay systems.}  Other  wireless networks such as cognitive radio networks (CRNs),  cellular, mesh, and  ad hoc have not been considered with \amj{various}  fading models.  

Further,  EC of multiple antenna systems (distributed and co-located) has been discussed in \cite{R315}. MIMO systems have been analyzed while \blue{the} establishing the EC perspective and  performance comparison of  distributed and co-located antenna systems. This study is a good contribution related to \blue{the} EC analysis of large scale MIMO systems. However, \cite{R315} does not cover in much detail \blue{the} EC-based performance evaluation of wireless communications. While  EC characterization of  \blue{multiple} antenna systems has been discussed  for mobile networks in \cite{R315}, \blue{the}  EC perspective of other  \blue{advanced} and emerging wireless networks has not been considered. On another note, \blue{the} EC-based performance evaluation of CRNs  while considering channel sensing and spectrum management paradigm in \blue{a} dynamic spectrum access environment has been discussed in \cite{R03} wherein an EC-based model for assessing the performance of CRNs with \amj{various}  PUs activity patterns has been developed.  \blue{While this work discusses the basis for analysing EC in CRNs, it only focuses on CRNs  \blue{with various}  spectrum sensing features.}

We note that an extensive work has   been done on  EC-based performance evaluation of   different communications networks but   \blue{a comprehensive survey is
yet to be conducted on this topic.}   This has motivated us to present a comprehensive survey of the cross-layer EC model that can be used to test the delay-limited  performance of various wireless networks. 

\subsection{Article Structure}
Table \ref{acronym} shows  different acronyms that has been used in the article. The rest of the article is organized as follows: Section \ref{sec:EC-defintions} presents the basic definitions of EC and effective-bandwidth. Section \ref{sec:case-studies} covers the five case studies (in five  networks) that employ the EC model, and in Section \ref{sec:applications}, \amj{various}  delay-constrained  applications \amj{for which their performance} is  tested through EC   have been surveyed.   Classification of \amj{various} fading models with \blue{their corresponding} achievable EC has been provided in Section \ref{sec:path-loss-models}, while in Section \ref{sec:networks}, different wireless networks with \blue{the  EC concept } has been discussed in detail. Section \ref{sec:full-duplex} discusses the  achievable EC in FD communications and Section \ref{sec:retransmission} covers the \amj{various} retransmission schemes that have been analyzed with \blue{the} EC concept.  Open issues and future research directions have been discussed in Section \ref{sec:open-issues}.   The entire article has been concluded in Section \ref{sec:conclusion}.

\section{Theory of Effective Capacity }
\label{sec:EC-defintions}
In this section, a basic  overview of EC has been provided. \amj{We note that EC is the dual concept of effective-bandwidth as presented in the pioneer work of Dapeng Wu in \cite{R110}. Hence, in order to have a good understanding of EC,} \blue{we first introduce the concept of effective bandwidth. }

\subsection{Effective Bandwidth}
\amj{The concept of effective-bandwidth is derived  through \blue{the}   large-deviation principle and can show  the   \blue{minimum}  constant service rate that is needed to satisfy a given queueing delay requirement for a given source rate  \cite{poor1}. Effective bandwidth  has been used extensively for obtaining optimal resource allocation schemes.}   \blue{Since} effective bandwidth is based on  large-deviation principle,  it is   \amj{traditionally used}  in  systems   \blue{ with large}  delay bounds.   However, effective  bandwidth has also been \amj{recently} used in  scenarios where  delay bound is short.    \blue{For example, in \cite{tony1}, the}  concept of effective bandwidth  \blue{is}  used to design  \blue{an adaptive}  resource allocation scheme \blue{for a system with ultra low latency requirements}.  Below is the description of the effective-bandwidth approximation:

Let us consider a  first-in-first-out (FIFO) queue with packets arrival rate at $t$ as $\mu(t)$,   \blue{the}  \amj{number of } packets in the queue  as  $q(t)$,  capacity of the link at time $t$ \amj{as} $c(t)$      \amj{and an}  infinite buffer size.   \amj{We consider}  $q(t)$ \amj{to} converge to a steady state  $q(\infty)$ and define   $A(t_{1},t_2)=\sum_{t=t_{1}+1}^{t_2}\mu(t)$ \blue{as the total number of arrivals at $(t_{1},t_2]$}  and $C(t_{1},t_2)=\sum_{t=t_{1}+1}^{t_2}c(t)$.

Authors in \cite{proof1,proof2} have proposed a  theorem to derive the theory of effective bandwidth. For this purpose, the following assumptions are used as presented in \amj{\cite{proof1,proof2}}.\\
Let assume
\begin{itemize}
 \item Arrival rate $\mu(t)$ and the service rate $c(t)$  are \blue{both} ergodic and stationary. Also $\mathbb{E}[\mu(t)]<\mathbb{E}[c(t)]$, where $\mathbb{E}[.]$ \blue{shows} the expectation operator.
 \item Arrival rate and source rate ($\mu(t)$ and $c(t)$) are  independent.
 \item Using the \amj{Gartner-Ellis}  theorem, we have  for all $\theta\in\mathbb{R}, \Lambda_{A}(\theta) \triangleq \textrm{lim}_{t\to\infty} \frac{1}{t} \textrm{log}(\mathbb{E}[e^{\theta A(0,t)}])$ and $\Lambda_{C}(\theta)\triangleq \textrm{lim}_{t\to\infty} \frac{1}{t} \textrm{log}(\mathbb{E}[e^{\theta C(0,t)}])$,  \amj{where $\theta$ is the delay exponent},  and  $\Lambda_{A}(\theta)$  and  $\Lambda_{C}(\theta)$ are assumed differentiable. 
 \end{itemize}
\blue{Now if} there exists a unique  $\theta^*> 0$,   \blue{such that  the below equation holds between the effective bandwidth and EC: } 
\begin{equation}
 \Lambda_{A}(\theta^*) + \Lambda_{C}(-\theta^*)=0,
 \label{eq2}
\end{equation}
\blue{then, mathematical derivations are provided in \cite{proof1,proof2} that relates the value of $\theta^*$ (found from (\ref{eq2}))  to the buffer overflow probability according to }
\begin{equation}
 \textrm{lim}_{x\to\infty}\frac{\textrm{log(Pr}\left \{  q(\infty)\geq x\right \} )}{x}=-\theta^*,
 \label{eq3}
 \end{equation}
\blue{where Pr$\{a\geq b\}$ shows the probability of $a$ being greater than or equal to  $b$.}
\\
\amj{Proof. For proof please refer to  \cite{proof1,proof2}.}
\\
Let $x$ be \blue{the}  buffer size of the link. Packets are usually dropped when    buffer becomes full. From  (\ref{eq3})  packet loss probability $\epsilon$ can be approximated as \cite{R323} 
\begin{equation}
\label{epsilon}
 \epsilon=e^{-\theta^*x}.
\end{equation}

\amj{Assuming that the  \blue{link} has  constant capacity such that $c(t)=c$, for all $t$,} \blue{$\Lambda_{C}(-\theta^*)$ can be simiplied as}

\begin{equation}
 \Lambda_{C}(-\theta^*)=\textrm{lim}_{t\to\infty}\frac{1}{t} \textrm{log}\left (  e^{-\theta^*ct}\right )=-\theta^*c.
\end{equation}
Using (\ref{eq2}), we  \blue{get}  $\frac{\Lambda_{A}(\theta^*)}{\theta^*}=c$. To have a small packet loss probability,  a capacity of the link equal to $\frac{\Lambda_{A}(\theta^*)}{\theta^*}$ is required \blue{where the value for $\theta^*$ comes from the} unique solution of $\theta^*=-(\textrm{log}\epsilon)/x$ (derived from \ref{epsilon}).   
 

  
\subsection{ Effective Capacity}
Authors in \cite{R110} introduced the concept of EC by taking  motivations from the theory of effective bandwidth.   \amj{EC  is the dual concept of effective bandwidth. Recall that  effective bandwidth shows the minimum service rate that is needed to guarantee a delay requirement for a given source traffic.   \blue{The  EC model}, on the other hand, can be used to find the maximum source rate that the channel can handle (service rate)  with the required delay constraint}.  As has been discussed above, \blue{the} concept of EC can be used when  \amj{a} delay bound is large. However, it can also be used to test the performance of  \amj{a} system  \amj{when}  delay bound is  \blue{small}, as has been discussed  \blue{in}  \cite{tony1}.    Analytical framework for deriving  EC has been discussed briefly below:

We assume the service process ${c(t),t=0,1,2,..,}$ with a partial sum $C(t1,t2)=\sum_{t=t_{1}+1}^{t_2}c(t)$ \blue{is} ergodic and stationary. Further, the  \amj{Gartner-Ellis}  limits for this service process is expressed as  
\begin{equation}
 \Lambda_{C}(\theta)\triangleq\textrm{lim}_{t\to\infty} \frac{1}{t} \textrm{log}(\mathbb{E}[e^{\theta C(0,t)}]).
\label{ec-1}
 \end{equation}
From (\ref{eq2}), we  get

\begin{equation}
 E_{c}(\theta^*)=-\frac{\Lambda_{C}(-\theta^*)}{\theta^*}=\mu.
 \label{ec-eq3}
\end{equation}
We recall that (\ref{ec-eq3}) is   EC of  service process, while $\theta^*$ is  the QoS exponent. The delay outage probability can now be formulated as

\begin{equation}
 \textrm{lim}_{x\to\infty}\frac{\textrm{log(Pr}\left \{  q(\infty)\geq x\right \} )}{x}=-\theta^*.
 \label{ec-eq322}
\end{equation}
A  more stringent QoS requirements can be represented by a larger value of   $\theta^*$ with a faster decay rate. However,  smaller values of $\theta^*$ represent   slower decay rates and provide  looser QoS guarantees.
\\
Now, an expression for the delay  $(D(t))$ experienced  by a packet at any time $t$ can also be approximated as follows
\begin{equation}
 \textrm{Pr}\left \{  D(t)>D_\textrm{max}\right \}\approx \textrm{Pr}\left \{  q(\infty)>0\right \}e^{-\theta^{*}\mu D_\textrm{max}},
 \label{ec-eq14}
\end{equation}
\amj{where  $\textrm{Pr}\left \{ q(\infty)>0 \right \}$ is the probability of non-empty buffer and $D_\textrm{max}$ is the \amj{maximum} delay bound. EC is the combination of two functions, namely,  QoS exponent and probability of non-empty buffer.}

\blue{The}  probability of non-empty buffer can be achieved by considering the 
\begin{equation}
\textrm{Pr}\left \{ q(\infty)>0 \right \} \approx  \frac{\mathbb{E}[\mu(t)]}{\mathbb{E}[c(t)]}.
 \label{ec-eq15}
\end{equation}

The above analytical explanation of effective-bandwidth and EC can be summarized as follows:
\begin{itemize}
  \item  \blue{The value of EC at $\theta^*$, $E_c(\theta^*),$       shows the maximum constant arrival rate.   Hence, $\mu \leq  E_c(\theta^*))$ should hold. }
 \item      The solution \blue{for $\theta^*$} can be found when  $E_{b}(\theta^*)=E_{c}(\theta^*)$ (for the arrival and source processes) holds. 
 \item Using  (\ref{ec-eq15}), the probability of non-empty buffer can be estimated. 
 \item Using  (\ref{ec-eq14}), the delay-violation probability can be estimated by using the pre-determined value of delay bound, probability of non-empty buffer, and obtained value of $\theta^*$. 
\end{itemize}

%

\section{Case Studies Involving Effective Capacity Measures}
\label{sec:case-studies}
%
%

To illustrate  \amj{an} in-depth understanding of \blue{the} EC metric and how it  \amj{can}  be used to  \amj{analyze}  the performance of    different network architectures or application scenarios, five  case studies have been presented. These include    device-to-device (D2D) communications, cellular networks, full-duplex communications, peer-to-peer communications, and visible light communications.  \amj{These}  case studies   show  \amj{the performance of}  various networks \amj{when handling delay-sensitive multimedia applications}  \amj{by using}  EC concept. The motivation of providing these case studies is to show  \blue{the} broad  \blue{use of the} EC modelling. \blue{In fact,} EC can be used to test the  \amj{performance} of  diverse network topologies, resource  \blue{allocation}  schemes,  traffic characterization, and various admission control policies.  Brief discussion of each case study has been described below:

\subsection{Cellular Communications}

EC-based delay analysis while considering  cellular communications has been well investigated in  literature \cite{R264,R105,R291,ner4,ner5}. QoS-aware real-time and delay-sensitive applications have been evaluated using \blue{the}  EC metric with different channel conditions and imperfections.   In \cite{R61},   the main architecture of cellular network involving \blue{the  EC model} with one mobile station (MS) and a base station (BS) is considered. In this  cellular network,  resources have been allocated based on  QoS-constraints  using \blue{the  EC model}.  A  queueing behavior at  data-link layer has been analysed  \amj{by investigating}   the \amj{maximum} achievable EC. \amj{The EC model has been used either at MS or BS to evaluate the performance of the network when handling a QoS-aware traffic.} \blue{ Moreover,   the QoS constraints are categorized either homogeneous or heterogeneous, depending on the known symmetrical and asymmetrical EC regions. This known EC regions show the impact of delay exponent in shaping the QoS constraints for two and three MS cases.}
\subsection{Peer-to-Peer Video Streaming}
\label{subsec:p2p-stream-case-study}
\amj{In peer-to-peer streaming, the EC model has been used to analyze the network performance.} \amj{In detail, to}   efficiently analyse the  peer-to-peer streaming, authors in \cite{R58} have  incorporated the concept of \blue{an} EC peer-selection (ECPS) approach for mobile users. In  the  \amj{proposed}  approach,  mobile users can enjoy  efficient video streaming without facing  \amj{long} delay.   In this peer-to-peer streaming, multiple attribute decision making  (MADM) approaches  have been used to accommodate various factors such as power-level, signal-to-interference and noise ratio (SINR) and mobility of peers.     

\subsection{Visible Light Communications}
Using the visible light, which is between 400 and 800 THz band, as  \amj{a} communication medium for  next generation wireless networks promises enhanced data rate for many delay-sensitive applications \cite{R117}. 

 Usage of multiple radio access schemes, such as millimeter-wave, UMTS, WLAN, and visible light communications can also result into  overlapping of their coverage area. This \amj{approach}  has resulted into the heterogeneous networks domain to cover various radio access technologies of next-generation networks. Most of the existing work on EC considers  visible light communication scenario    \amj{within the}  heterogeneous network architecture with cellular networks (such as femto cells as in \cite{R118}).  Existing works on visible light communications usually take into consideration \blue{the} EC concept to assess   QoS-awareness in heterogeneous networks  \cite{R135}. With the introduction of   \blue{the  EC model} in the optical communications,  satisfying the  statistical delay requirements    while supporting  user-centric (UC) cluster formation \blue{has been investigated}. \blue{The achievable EC and the sum utility functions are quantified for the UC cluster-formation. By applying the concept of EC and sum utility, the problem becomes tractable, and is solved by using the exhaustive search.  This analysis   shows that the UC   cluster formation  achieves  higher  EC  compared to conventional cellular network designs. }                  

\subsection{Full Duplex Communications}
With the advances in self-interference suppression (SIS) approaches, the dream of full-duplex (FD)  communications has been realized\footnote{for detail on FD and SIS approaches see Section \ref{sec:full-duplex}.}. With the advent of FD communications,  simultaneous sending and receiving on the same spectral band can almost double the throughput as compared to  half-duplex communications \cite{R21}. With enhanced data rate,  FD communications has been extensively studied for  multimedia applications with stringent delay requirements \cite{R221}.  \blue{The  EC model}  has also been \amj{used in}  FD communications to test  \amj{the performance of the network for}  various QoS-aware applications \cite{R218}. However, \amj{the maximum achievable EC of}   \amj{many}  FD paradigms, such as, directionality, beamforming, and various transmit \blue{and} receive antenna pairing \amj{schemes is not known yet}.   Most of the work on FD communications with consideration of EC has been presented with  FD-relay networks.  \amj{For example, the}  maximum constant arrival rate  \amj{of an}  FD-enabled  \blue{communication system}  \amj{while satisfying a predefined delay constraint is found in \cite{R221}}. With the implementation of proper SIS approach, such as, passive or active SIS approach,   \amj{the maximum achievable EC of an} FD communication paradigm has been found in \cite{R309}. \blue{In this study, by finding the optimum fixed value for the source arrival rate (EC), the properties of the source and relay has been investiaged. Afterwards, the depending on the achievable EC of the source, the optimal resource allocation for the relay and source are derived. }        

\subsection{Summary and Insights}
In this section, five case studies encompassing the concept of EC has been showcased to understand a wide-range of applicability of the EC model ranging from cellular networks to visible light communications. \amj{Ensuring the QoS constraints in wireless networks that deal with delay-sensitive applications is \amj{in fact}  a challenging task.  We note that, channel conditions determine the capacity of a network, and as such, whether or not the required QoS constraints is achievable. The variability in wireless channels results in variability in the transmission buffer status and, in turn, \blue{in} the experienced delay by the transmitted packets. The concept of EC, that  \blue{takes into consideration }  the physical layer parameters  \blue{in conjunction with}  the link-layer and provides a simple formulation for a link-layer performance is well received by the researchers working in various networks. This concept, not only can be used for analyzing the performance of the networks, but it also provides a strong mathematical framework for efficient design of the system parameters.   This section is clear description of a wide applicability of \blue{the  EC model}  in traditional, as well as, emerging  wireless networks such as  \blue{URLLC and}   visible light communications.} 
\begin{table*}
\scriptsize
\centering
\caption{Different Delay-sensitive applications Involving  Effective Capacity Measurements}
\label{tab:wmcrns-applications}
\begin{tabular}{|p{3cm}|p{3cm}|p{3.5cm}|p{1cm}|p{3.5cm}|}
\hline

 \multicolumn{1}{|c}{\bf Different Applications} &   &   \bf  Architecture or Network Used&\bf Papers&\bf Fading Channel Used     \\\hline
 Voice Applications   &  Voice Calls  & Cellular Network &\cite{R105} &Rician Fading Channels   \\\cline{3-5} 
      &  & OFDMA-Based Networks & \cite{R226}&Rayleigh Fading Channels   \\\cline{3-5} 
 
  &  & Multi-hop Networks & \cite{R235}& Not Defined  \\\cline{3-5} 
 
  &  &Cognitive Radio Networks  & \cite{R243}&Rayleigh Fading Channels   \\\cline{3-5} 
 
  &  & Multi-hop Networks & \cite{R246}&  Not Defined \\\cline{3-5} 
 
  &  & Cross-layer Network Design & \cite{R270}& Nakagami-$m$  Fading Channels  \\\cline{3-5} 
 
  &  & Proactive  \amj{Link Selection Routing}  & \cite{R278}& Not Defined  \\\cline{2-5}
 & VoIP Applications  & Long Term Evolution  &\cite{R176}   &Rayleigh  Fading Channels \\\cline{3-5}
 &  & Multi-user Network Layout  & \cite{R220}  &Rician Fading Channels \\\cline{3-5}
 &  &Wireless Sensor Networks   &   \cite{R228}&Rayleigh  Fading Channels\\\cline{3-5}
 &  & Cognitive Radio Networks  & \cite{R229}  &Rayleigh Fading Channels\\\cline{3-5}
 &  & Relay Networks  &  \cite{R231} &Rician Fading Channels \\\cline{2-5}
           &Cellular Telephony & Cellular Networks  &\cite{R110}   &Rayleigh Fading Channels  \\\hline
    Miscellaneous applications   &Medical Application &WiMAX Networks   &\cite{R193}   & Not Defined \\\cline{2-5} 
   & Smart Grid Application &Non-Intrusive Application   &\cite{R19}   &Rician Fading Channels  \\\cline{2-5}
   &Image Transmission  & Multi-hop Mesh Networks & \cite{R266}&Rayleigh Fading Channels  \\\hline
 Video Applications &Video Streaming  &Mobile Networks  & \cite{R270}&Nakagami-$m$  Fading Channels \\\cline{3-5}
 & &Cross-layer Design  & \cite{R10}&Rayleigh Fading Channels  \\\cline{3-5}
&  & Long Term Evolution & \cite{R33}& EPA Fading Channels  \\\cline{3-5}
 & & Cross-layer Design & \cite{R47}&Rayleigh Fading Channels  \\\cline{3-5}
  & &Wireless Cooperative Networks  & \cite{R56}&Generalized $k$  Fading Channels   \\\cline{3-5}
   & &Multi-user Video Streaming  & \cite{R86}&Correlated  Fading Channels   \\\cline{3-5}
    & &Multi-Channel  & \cite{R89}&Rayleigh Fading Channels   \\\cline{3-5}
     & &Broadband-ISDN   & \cite{R113}&Not Considered   \\\cline{3-5}
      & &Multi-user Video Streaming  & \cite{R119}&Rayleigh Fading Channels   \\\cline{3-5}
       & & Wireless Local Area Networks & \cite{R121}&Nakagami-$m$  Fading Channels  \\\cline{3-5}
        & &5G Networks  & \cite{R122}&Nakagami-$m$ Fading Channels   \\\cline{3-5}
         & &FD-Relay Networks  & \cite{R123}&Rayleigh Fading Channels  \\\cline{3-5}
          & &Cognitive Radio Networks  & \cite{R124}&Nakagami-$m$ Fading Channels  \\\cline{3-5}
           & &5G Networks  & \cite{R125}&Nakagami-$m$ Fading Channels   \\\cline{3-5}
                      & & FD-Relay Networks & \cite{R133}&Nakagami-$m$ Fading Channels   \\\cline{3-5}
                      & & Femto Cells & \cite{R140}& Not Considered  \\\cline{3-5}
                                           
                      & & Wireless Local Area Networks & \cite{R146}& DTMC-Based Fading Channels   \\\cline{3-5}
                                          
                      & &OFDMA-Based Networks  & \cite{R164}&Nakagami-$m$ Fading Channels   \\\cline{3-5}
                                        
                      & & Wireless Local Area Networks & \cite{R173}&Rayleigh Fading Channels   \\\cline{3-5}
                                           
                      & &Multi-User Video Streaming  & \cite{R188}&Rayleigh Fading Channels  \\\cline{3-5}
                                        
                      & &Multi-User Video Streaming  & \cite{R194}&Rayleigh Fading Channels  \\\cline{3-5}
                                         
                      & &Cross-Layer Design  & \cite{R204}&Rayleigh Fading Channels  \\\cline{3-5}
                      & & WiMAX & \cite{R206}&Rician Fading Channels  \\\cline{3-5}
                      & &Multi-User Video Streaming  & \cite{R212}& Not Defined  \\\cline{3-5}
                                           
                      & &Wireless Virtual Networks  & \cite{R223}&Rayleigh Fading Channels  \\\cline{3-5}
                                          
                      & &OFDMA-Based Networks  & \cite{R237}&Rayleigh Fading Channels  \\\cline{3-5}
                                        
                      & &Wireless Sensor Networks  & \cite{R239}&Rayleigh Fading Channels   \\\cline{3-5}
                                           
                      & &Multi-User Video Streaming  & \cite{R244}&Rayleigh Fading Channels  \\\cline{3-5}
                                        
                      & &Wireless Local Area Networks  & \cite{R260}& Not Defined  \\\cline{3-5}
                                         
                      & & Cross-Layer Network Design & \cite{R268}&Nakagami-$m$  Fading Channels  \\\cline{3-5}                                         
                       & &Heterogeneous Wireless Networks  & \cite{R277}&Not Defined   \\\cline{3-5}
                      & &Cellular Networks  & \cite{R281}&Rayleigh Fading Channels  \\\cline{3-5}
                                           
                      & &Single User Video application  & \cite{R288}&Rayleigh Fading Channels  \\\cline{2-5}
                                          
                      &High Speed Video Transmission & Secure Wireless Networks & \cite{R174}&Nakagami-$m$ Fading Channels  \\\cline{3-5}
                                        
                      & &Multi-User Video Transmission  & \cite{R205}&DTMC-Based Fading Channels  \\\cline{3-5}
                                           
                      & &Two-Hop Networks  & \cite{R222}&Rayleigh Fading Channels  \\\cline{2-5}
                                        
                      & Light Video Transmission&Secure Wireless Networks  & \cite{R174}&Nakagami-$m$ Fading Channels  \\\cline{3-5}
                                         
                      & &Multi-user Video Transmission  & \cite{R205}&DTMC-Based Fading Channels   \\\cline{3-5}
                      & &Two-Hop Networks  & \cite{R222}&Rayleigh Fading Channels  \\\cline{2-5}
                      &Peer-to-Peer Streaming &Peer-to-Peer Networks  & \cite{R58}&Not Defined   \\\cline{2-5}
                      &Video Conferencing & Multi-User Video Transmission & \cite{R98}&Rayleigh Fading Channels  \\\hline
\end{tabular}
\label{table:all-applications} 
\end{table*}

\section{\amj{ Various QoS-aware Applications Analysed With the EC Model}}
\label{sec:applications}
In this section, we  provide an in-depth analysis of various applications  \blue{using the EC model}.  Table \ref{table:all-applications} presents the details of various applications which are specifically analysed with the concept of EC in various networks. Different applications have been classified into voice, video, and miscellaneous applications. This classification is based on the delay requirements and network architecture used with respect to different applications. 
\subsection{Voice Applications}
\label{subsec:voice-application}
Plethora of applications can be tested with EC, while taking into consideration  certain QoS requirements  \cite{R153}. We have further classified the voice applications into voice call, VoIP, and cellular telephony. \amj{Different voice applications are classified based on  different communication paradigm used such as voice over Internet protocol or traditional wireless networks.} Details of each sub-category have been presented below:

\subsubsection{Voice Calls}
 Idea of a link-layer channel model with QoS-aware metric support  has been used in various networks with different architectures to assess the quality  \amj{of} voice calls. Compared to  video calls, voice calls are regarded as  delay-sensitive low data rate applications \cite{R226}. To analyse the quality of voice calls with small delay, EC-based source traffic and service characterization can be carried out. In \cite{R105},   the \amj{EC model has been used to analyse the performance of the proposed, \blue{but} not necessarily perfect, design methods to support voice applications.}     Four antennas  \blue{have}  been used, and achievable EC  \blue{was obtained} to test the  delay-limited  performance \blue{of multiple antenna systems} for voice applications. \blue{This multi-antenna based EC analysis shows that 90\% of the system performance or voice quality is reduced due to  imperfections in the system. This framework was also used to investigate the enabling technologies such as W-WCDMA and MC CDMA for the fourth generation wireless networks.}    Various quality constraints for  voice calls with  EC as the performance metric  have been investigated in \cite{R235,R246}. In this study,  issue of service degradation and source dissatisfaction has been tested  \amj{using}  EC metric while  \amj{proposing an}  optimal resource allocation  scheme. 

An optimal resource allocation scheme,  \amj{using the EC model,}  for  voice calls in CRNs has been presented in \cite{R243}.  \amj{Using}  EC,  \amj{an} optimal sensing time and channel allocation  \amj{scheme}    has been analysed in detail. The proposed resource allocation scheme has also been evaluated through extensive simulations to show its effectiveness. Cross-layer EC modeling for  testing performance of QoS-aware applications such as   independent and identically distributed (i.i.d.) and  non i.i.d fading channels were estimated while taking into consideration the multicast receivers \blue{was studied in \cite{R270}.}  Extensive simulations were also performed to clearly demonstrate the trade-off between different QoS metrics.  \blue{The} EC   \blue{model} to test the quality of  voice and other delay applications has also been presented with   unicast routing control agent (URCA) \amj{in} \cite{R278}.  The proposed routing scheme can evenly distribute the load over all the available paths,   \amj{and} minimize the impact of link failure on the performance of  network. \blue{From the achieveable EC of the individual links, the soft link failures are predicted. The soft link failures are then minimized to ensure the success of critical voice sessions. }  

\subsubsection{VoIP Applications}
With the advancement in Internet protocol (IP)-based networking,  voice over Internet protocol (VoIP) applications have gained  \blue{a lot of attention}  . \amj{The maximum achievable EC of}  VoIP or IP telephony has been extensively studied and analysed    while considering various network and architectural designs under different fading channels in \cite{R176}.  In \cite{R220},  achievable EC of  wireless networks with multiple input single output (MISO) \amj{for VoIP applications} has been investigated in detail. As compared to the traditional work that   considers   Rayleigh fading, this work considers the provision of \amj{statistical}  QoS-guarantee under Rician fading channels. With the help of EC concept,  effective rate that can support  future applications like VoIP,  has been measured .  A routing protocol that takes into consideration the end-to-end delay  \amj{for wireless sensor networks}   has been proposed in \cite{R228}. \amj{This routing scheme is then investigated with the help of \blue{the} EC concept to find the shortest possible paths while residing within the required delay constraints. This routing scheme is then tested with the VoIP applications.}

In addition to WSNs,  VoIP applications have also been investigated in CRNs with the concept of EC  \cite{R229}.  \blue{In particular, an EC-based }  optimal resource  \blue{allocation scheme for}  cognitive radio (CR)-based femto cells have been \amj{investigated.}  This  resource allocation scheme   \amj{also takes into consideration } the cross-tier interference, and \amj{hence}, provides a significant support for  delay-sensitive applications such as VoIP. VoIP applications support in emerging futuristic networks such as mobile multi-hop networks with the  EC \amj{model}   has also been discussed in \cite{R231}. In this study, authors have investigated  \amj{the achievable EC of multi-hop mobile networks and then assess the network functionality with different delay-bounds.} \amj{In this work,} a cross-layer simulation platform has also been proposed to study the impact of various delay-sensitive applications such as VoIP \amj{on} multi-hop  network model.    
\subsubsection{Cellular Telephony}
Simplest of the voice applications is the traditional cellular telephony. In \cite{R110}, authors have used the concept of EC \amj{to investigate the QoS in for a simple scenario of cellular telephony.}   In this study, a comparative view of physical layer channel model and link-layer channel model (EC)  has been investigated.  Cellular telephony has been tested by  physical as well as by  \amj{EC} parameters.   With the help of two EC functions, a wireless link has been modeled to provide the QoS-guarantee for the delay-sensitive traffic. \blue{This study is the pioneering work on the EC concept based on the idea of effective bandwidth . This was the first attempt to investigate the link-layer channel model while taking into consideration the statistical QoS provisioning in wireless networks.}        

\subsection{Miscellaneous Applications}
\label{subsec:misc}
  Existing work on  \blue{the  EC model}  in wireless networks can also take into consideration other   \amj{low latency} and real-time applications such as 2D, monitoring, medical, and smart grid applications  \cite{R266,R296,R297}. Below is the description of  various    applications  \amj{that require low latency and} that have been discussed  \amj{in conjunction with the EC concept in}   wireless networks.

\subsubsection{Medical Applications}
 \blue{The}  EC \blue{model} has  been  \blue{used}  to test \blue{the performance of the system for} certain advance medical applications. In \cite{R193}, authors have  \amj{investigated}  end-to-end delay distributions for  tele-ultrasonography based on the EC modeling of wireless channels. For this purpose, a cross-layer simulation platform that consists of a source of medical ultrasound  at a remote location has been analyzed  over the WiMAX link.   Extensive simulations show the effectiveness of the proposed scheme.    

\subsubsection{Smart Grid Applications}
 Traditional power grid has been transformed into the intelligent smart grid \cite{Rehmani17tii,Rehmani16adhocnets}.  Smart grid   generates different types of multimedia traffic and  \amj{CR}  is considered to be a potential technology which can assist  multimedia applications in smart grid environment~\cite{Wang13wc,Khan16comst}.  In \cite{R19},  authors have used  \blue{the  EC model} to measure  quantitatively the support of various smart grid multimedia applications in existing wireless communications designs. Different case studies considering various smart grid multimedia applications with their implementation in various communication scenarios with   \blue{the  EC model}  has been discussed in detail. \blue{In this case, EC comes out to be an efficient tool to quantify the performance of  various smart grid applications regardless of  different network technologies.}

\subsection{Video Applications}
\label{subsec:videoapplications}
An extensive work on  achievable EC in wireless networks considers  video applications as the test applications to assess     \amj{the various delay requirements of a system}  \cite{R58,R86,R89,R113}. \amj{In this classification, most of the studies consider  delay-sensitive video applications using different video codecs. In comparison to the voice applications, various network designs with video transmission capabilities have been investigated with their achievable EC.}  

Video applications are regarded as  delay-sensitive and time-critical application, that require  QoS-guarantees. In this subsection,   \amj{ an overview of} various video applications such as video streaming and video conferencing  has been   \amj{presented} with  \blue{the  EC model}.

\subsubsection{Video Streaming}
 Video streaming is a challenging application due to its strict delay bound  and bursty flow. With these limitations,  transmission of videos over  wireless medium  \amj{with stringent delay requirements} seems to be  \amj{a challenging task}  \cite{R223}.   \blue{In this respect, the EC model can be used}    \amj{for assessing} the performance of various \amj{low latency} applications with stringent QoS requirement \amj{while residing within a given delay violation probability.}

 \blue{Indeed an}  extensive work has been done on EC \amj{modelling of wireless channel} while taking into account the  video streaming applications \cite{R270,R10,R33,R47,R56,R119,R121,R122,R123,R124,R125,R133,R140,R146,R164,R173,R188,R194,R204,R206,R212,R237,R239,R244,R260,R268,R277,R281,R288}. In  the above mentioned works,  authors have used the  EC model  \amj{in}  different fading conditions, such as Rayleigh, Nakagami-$m$, and Rician. However, most of the works of \blue{the  EC model} with video related applications consider  Rayleigh fading channels. Only the works in \cite{R121,R270,R124,R125,R133,R164,R268} consider  Nakagami-$m$ fading channels with \blue{the  EC model}. Video streaming support have \amj{also} been \amj{investigated} with EC concept while using  Rician fading channels \cite{R206}. \blue{In this study, the achievable EC has been studied with the physical layer IEEE 802.16-2004 WiMAX standard. By using the video streaming scenario with different channel conditions, different delay values are estimated to gaurantee the uninterrupted video quality.}

\amj{Video surveillance or monitoring applications were also investigated with \blue{the} EC \blue{model}  in \cite{R159}. In this work,  concept of EC has been used to forsee the QoS guarantee in cognitive relay networks. \blue{In this work, the}  performance of a cognitive relay (COR) algorithm to  support  efficient video surveillance has been  \amj{analyzed}  in terms of its  \amj{achievable} EC.  This COR approach    has also been \amj{explored} for  machine-to-machine (M2M) communications.  }

\subsubsection{High/Low Speed Video Transmission}
\amj{To simplify the classification of different video applications depending on the delay requirements, we can also categorize them into high and low speed. For example, the video-conferencing and online gaming with stringent delay requirements can be categorized as high speed video transmission. Other applications such as traditional video-streaming with less stringent delay and date rate requirements can be classified as low-speed video transmission.}

High  speed video transmission demands more control and resources as compared to the traditional video streaming applications \cite{R174}.  Authors in \cite{R222}, have provided a QoS-aware power allocation based on achievable EC  over  two hop networks. This resource allocation scheme is then utilized to support high and low speed video transmission.  Total power consumption is also minimized \amj{while} guaranteeing the specific QoS requirements. \blue{In comparison to  single hop networks, this study focuses on two hop DF relaying transmissions. Using the EC metric,  delay distributions of  two hops are obtained. In order to provide  statistical-delay gaurantees for  two hops relays, it is estimated that the delay distributions of  both hops should be same, which then is achieved with the help of an asymmetric resource allocation scheme.}

\subsubsection{Video Conferencing}
Performance of  advanced real-time applications such as video conferencing has also been tested with  EC metric. In \cite{R98}, authors have  \blue{obtained}   EC  of multiple antenna systems  \blue{which} shows a significant \blue{EC} gain  \blue{using}   multiple antennas. Specifically, the achievable EC   \blue{of} a multi-antenna  Rayleigh fading channel with a procedure called channel hardening has been \blue{found}.   \blue{The} overall gain achieved  \blue{by using the EC model} has been exploited to achieve  smooth video conferencing without  \blue{a large}  delay.  

\subsection{Summary and Insights}
In this section, \blue{the} performance of  various real-time and delay-sensitive applications \blue{using} EC  \blue{model} has been explored for  different  \blue{applications}  such as various design, and architecture. Delay limited performance  analysis of QoS-aware applications shows that   data loss  \blue{could} occur when  delay thresholds are violated and  testing the performance of applications with stringent delay requirements \blue{are specifically more challenging}.  \blue{Indeed, it is a well-known fact that} for applications with stringent-delay and ultra-reliability requirements, such as online gaming, video-conferencing, and autonomous vehicles, using the physical layer only parameters for capacity estimation may not be accurate.   \blue{The} suitability of this mathematical framework to model the performance of the network  while residing within the stringent delay and reliability requirements \blue{need further research to be verified.}     
 
 \section{Effective Capacity analysis with different Fading models}
\label{sec:path-loss-models}

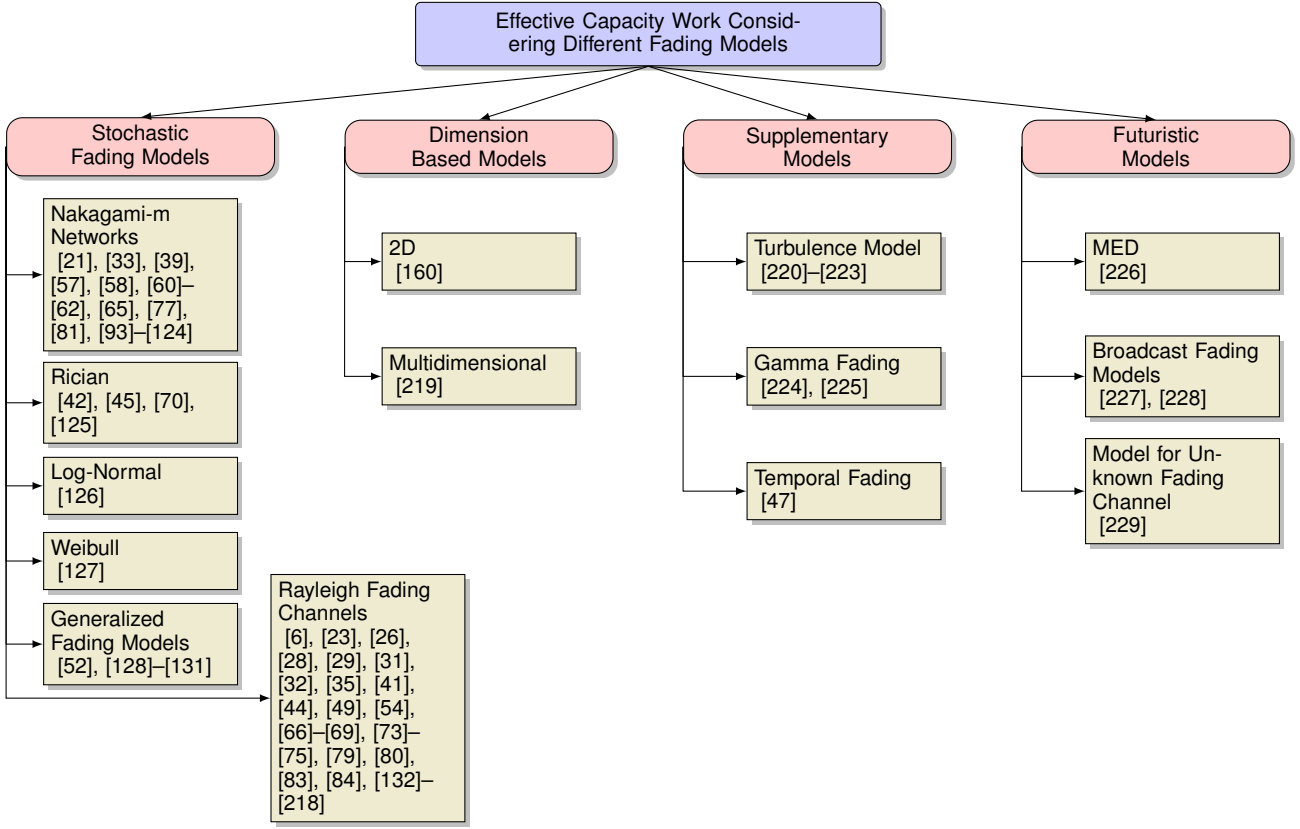
\begin {figure*}
\footnotesize
\centering
\begin{tikzpicture}[
  level 1/.style={sibling distance=45mm},
  edge from parent/.style={->,draw},
  >=latex]

\node[root,minimum height=3em,text width=6cm,fill=blue!20] {Effective Capacity Work Considering Different Fading  Models}
  child {node[level 2] (c1) {Stochastic \\ Fading Models    }}
  child {node[level 2] (c2) {Dimension\\Based Models   }}
  child {node[level 2] (c3) {Supplementary\\ Models }}
  child {node[level 2] (c4) {Futuristic \\Models  }};

\begin{scope}[every node/.style={level 3}]

\node [below of = c1,xshift=0pt,yshift=-20pt]	 (c11) {Nakagami-m  Networks\\\cite{R05,R12,R13,R14,R27,R41,R46,R53,R69,R82,R88,R94,R99,R121,R122,R124,R125,R129,R133,R134,R138,R139,R150,R157,R161,R164,R174,R182,R187,R202,R216,R218,R219,R236,R250,R253,R262,R264,R268,R270,R271,R282,R303}
};
\node [below of = c11,xshift=0pt,yshift=-20pt] (c12) {Rician\\\cite{R106,R206,R220,R231}};
\node [below of = c12,,xshift=0pt,yshift=-3pt] (c13) { Log-Normal\\  \cite{R103}};
\node [below of = c13,,xshift=-0pt,yshift=-0pt] (c14) { Weibull\\ \cite{R116}}; 
\node [below of = c14,,xshift=0pt,yshift=-3pt] (c15) { \amj{ Generalized \\Fading Models \\ \cite{R17,R20,R24,R195,R56}}};
\node [below of = c15, xshift=86pt,yshift=8pt] (c16) {Rayleigh Fading Channels\\
 \cite{R01,R02,R09,R10,R15,R18,R21,R23,R25,R26,R28,R31,R32,R34,R40,R42,R48,R51,R55,R61,R63,R65,R72,R73,R76,R80,R83,R84,R85,R87,R89,R90,R91,R93,R98,R110,R115,R117,R118,R120,R141,R143,R145,R149,R151,R154,R156,R160,R162,R165,R166,R167,R170,R171,R172,R173,R176,R177,R179,R180,R183,R184,R185,R186,R188,R190,R192,R194,R196,R199,R200,R201,R204,R207,R209,R210,R213,R215,R217,R221,R222,R225,R226,R229,R237,R239,R244,R245,R247,R255,R256,R258,R265,R267,R269,R272,R273,R274,R275,R276,R279,R280,R281,R283,R286,R287,R288,R289,R291,R298}
};
\node [below of = c2, xshift=00pt,yshift=-15pt] (c21) {2D\\\cite{R91} };
\node [below of = c21, yshift=-15pt] (c22) {Multidimensional\\\cite{R95} };
\node [below of = c3, xshift=10pt,yshift=-15pt] (c31) {Turbulence Model\\\cite{R07,R131,R147,R148} };
\node [below of = c31,yshift=-15pt] (c32) {Gamma Fading \\\cite{R16,R54}};
\node [below of = c32,yshift=-15pt] (c33) {Temporal Fading\\\cite{R19} };
\node [below of = c4, xshift=10pt,yshift=-15pt] (c41) {MED\\\cite{R06} };
\node [below of = c41,yshift=-15pt] (c42) {Broadcast Fading Models \\\cite{R96,R252}};
\node [below of = c42,yshift=-15pt] (c43) {Model for Unknown Fading Channel\\\cite{R155} };

\end{scope}

\foreach \value in {1,2,3,4,5,6}
  \draw[->] (c1.180) |- (c1\value.west);

\foreach \value in {1,...,2}
  \draw[->] (c2.180) |- (c2\value.west);

\foreach \value in {1,...,3}
   \draw[->] (c3.180) |- (c3\value.west);
   
\foreach \value in {1,2,3}
  \draw[->] (c4.180) |- (c4\value.west);

\end{tikzpicture}
\caption {\amj{The existing work on effective capacity with respect to different fading models can be classified in stochastic, generalized, futuristic, dimension-based, and supplementary fading models }}
\label{fig:pathlossmodels}
\end{figure*}
In this section, we have provided  \blue{a survey of existing work with their}   achievable EC with  different fading  models  \amj{used in various }    wireless networks. Figure \ref{fig:pathlossmodels} show  different fading models \amj{that have been taken into consideration}  with \blue{the} EC \amj{concept.}     We note that \amj{channel variability can cause    long delays   in the transmission buffers. Hence, indicating the importance of using a suitable mathematical framework for testing the performance of the networks.}  The EC model can \blue{indeed} be used  in designing the adaptive resource allocation and scheduling schemes \cite{R77} \blue{that are specifically suitable for applications with constraints on the buffer size}. The main advantages of utilizing  \blue{the  EC model} with different fading models  \blue{are the}   provision of \blue{a} general \amj{mathematical} framework \amj{and simplification of  QoS-aware metrics}.  
\subsection{Stochastic Fading Models}
Stochastic fading models  cover the fading in  \amj{a} channel that results from  scattering and multipath propagation.   \blue{In these models,}  a random variable is added  to show the additional fading.   

 We recall that EC provides a generalized  link-layer  \amj{mathematical} framework (its complexity has been discussed in Section \ref{subsec:fading_summary}) \blue{for testing the performance of the channel under delay constraints}.  On that basis,  different fading models can be analyzed with their distinct characters.   Existing work  \amj{on} \blue{the  EC model} mostly take into consideration the stochastic fading  \amj{models}  for   \amj{an} assessment  of QoS-awareness in \amj{wireless networks.}   Among the  stochastic fading models, Rayleigh and Nakagami-$m$ fading channels have been   \amj{used} extensively with  EC  \amj{concept.} Current work in wireless networks considers  different versions of stochastic fading models including Rayleigh, Nakagami-$m$, Rician, log-normal, and Weibull fading channels  with EC metric. Below is the description of each fading channel:

\subsubsection{Rayleigh Fading Channels}
Most of the existing work on EC in wireless communications considers Rayleigh fading channels. Rayleigh fading is more prominent when there is no line of sight communications between the transmitter and  receiver. Following works  \cite{R01,R02,R09,R10,R15,R18,R21,R23,R25,R26,R28,R31,R32,R34,R40,R42,R48,R51,R55,R61,R63,R65,R72,R73,R76,R80,R83,R84,R85,R87,R89,R90,R91,R93,R98,R110,R115,R117,R118,R120,R141,R143,R145,R149,R151,R154,R156,R160,R162,R165,R166,R167,R170,R171,R172,R173,R176,R177,R179,R180,R183,R184,R185,R186,R188,R190,R192,R194,R196,R199,R200,R201,R204,R207,R209,R210,R213,R215,R217,R221,R222,R225,R226,R229,R237,R239,R244,R245,R247,R251,R255,R256,R258,R265,R267,R269,R272,R273,R274,R275,R276,R279,R280,R281,R283,R286,R287,R288,R289,R291,R298} 
consider  Rayleigh fading  with \blue{the  EC model} in different wireless networks.  More prominent \amj{wireless} networks that  \amj{have been investigated with}   Rayleigh fading channels with \blue{the  EC model},  are CRNs, cellular networks, and  \amj{cooperative}  networks including the FD-relay networks.  In cellular networks, with statistical QoS provisioning, Rayleigh fading has been extensively evaluated with EC metric. In CRNs with multiple channels,  prediction related to  multiple interference has also been studied with  Rayleigh fading channels. Achievable EC in CRNs with multiple channels and Rayleigh fading as the physical channel model has been extensively investigated to  \amj{find the maximum arrival/source rate with the required delay-outage probability.}

\blue{Most of the delay-sensitive applications with Rayleigh fading in wireless networks  have also  been evaluated with  EC metric.} \amj{Rayleigh fading has been used extensively because it helps the researchers to understand the radio signals in heavily urban environment.  Closed-form expression of achievable EC with Rayleigh-fading is less complex as compared to the Nakagami-$m$ fading channel. Therefore, maximization in EC of different wireless networks with Rayleigh-fading has been investigated extensively in the existing works. } Another fading channel, that has been used extensively after  Rayleigh fading is  Nakagami-$m$ fading channel.

\subsubsection{Nakagami-m}
For EC-based delay \blue{performance} estimation  \blue{of}  wireless networks, where the large delay-time spreads are going to be estimated,  Nakagami-$m$ fading channels are used by clustering different reflected ways. Authors in  \cite{R05,R12,R13,R14,R27,R41,R46,R53,R69,R82,R88,R94,R99,R121,R122,R124,R125,R129,R133,R134,R138,R139,R150,R157,R161,R164,R174,R182,R187,R202,R216,R218,R219,R236,R250,R253,R262,R264,R268,R270,R271,R282,R303}, have considered  Nakagami-$m$ fading distributions   in different wireless networks using \blue{the  EC model}.  \amj{Nakagami-$m$ channel model is often regarded as the general fading channel and can be used to investigate the mobile and indoor-mobile scenarios. Depending upon the parameter $m$, where $m\in\left \{\frac{1}{2},+\infty  \right \}$, different fading conditions can be achieved. For example, when $m=\frac{1}{2}$, this represents the severe fading case, while $m=1$ is the Rayleigh-fading,  $m>1$ approximates the Rician channel,   and $m=\infty$ represents additive white Gaussian noise (AWGN).}

The main advantage of using  Nakagami-$m$ fading distribution with \blue{the  EC model} in wireless networks is better matching of its empirical data as compared to  other distributions such as Rayleigh and Rician. Authors in \cite{R157}, have investigated  \blue{the  EC model}  with Nakagami-$m$ fading channel. This study reveals that  uncorrelated Nakagami-$m$ flat fading conditions can well be analyzed with  EC-based QoS-aware model.  \blue{Complementary cumulative distribution function (CCDF) of
delay has also been approximated by the EC model in this work. This analytical approximation based on EC leads to understanding the delay statistical behavior, which is not possible with the physical layer channel models.  }

\amj{Works of} CRNs \cite{R202,R41,R88} and relay networks \cite{R53,R157} show that queueing behavior can well be evaluated with EC metric under Nakagami-$m$ fading conditions. 
\subsubsection{Rician}
As compared to  Rayleigh fading channel, in Rician fading, out of several different paths one  must be \amj{the} line of sight path. In this fading conditions,  amplitude of the propagated signals are modelled by using  Rician distributions.      \amj{Achievable EC of wireless networks with Rician fading conditions has been discussed in       \cite{R106,R206,R220,R231}. } As Rician fading conditions consider one strong component, this strong component can be the  line of sight path, therefore   Rician fading  can be  \amj{employed}  in   \blue{some advance }  networks such as satellite communications \cite{R74} \blue{which studies the EC model of channel. We note that the satellite communications suffer from long delays in signal transmission due to the very long distance between the satellite and the users. Hence, the concept of EC can be very useful for analyzing the performance of these communications systems. }   In addition to  satellite communications,  \blue{the  EC model}  with  Rician fading    \blue{has also been studied}  in cellular communications, indoor networks, and vehicular networks.

\subsubsection{Log-Normal}
In addition to  Rayleigh, Nakagami-$m$, and Rician fading conditions, other stochastic fading models  such as log-normal has also been  investigated with \blue{the  EC model}. In log-normal fading channel, the mean and distributions of  path loss signals that are treated as a random variable can be used to model the physical-layer wireless channel model. Authors in  \cite{R103},    \amj{have employed  EC to estimate the delay-outage probability with log-normal fading distribution. In this work, the system performance of the CDMA networks has been  \blue{investigated}  \blue{in detail}.  Network traffic has been modelled as a stochastic process, \blue{and then,} extensive simulations have been performed to show the impact of network traffic on the achievable EC of the system with log-normal fading conditions.} \blue{With this framework, the system capacity and  traffic demands are predicted. Also, this EC-based anlsyis leads to the conclusion that, traffic correlation is good when the system load level is the same.}  

\subsubsection{Weibull}
 This fading channel has been used in wireless communications with its implementation in indoor and outdoor environment.  \blue{The  EC model}  in   \amj{wireless}   networks with Weibull fading channel has been discussed in \cite{R116}. To support  real-time applications, an independent but not identical Weibull fading channel has been used to find and test the effective rate while considering the concept of EC. High SNR and low SNR-based closed-form asymptotic analysis has also been performed to find the optimal effective rate under \amj{other}  fading conditions  \amj{as well}.   \blue{This work shows that,  more stringent delay requirements minimizes the effective rate. Another very interesting conclusion from this work is that, the effective rate is prone to the severity of  fading channels. }
\subsubsection{Generalized Fading Models}
\amj{Compared to other fading models such as Rayleigh and Rician, generalized fading models provide a general framework with the combination of one or more fading model   \cite{R37}. In general,  EC concept can be used to analyze the performance of generalized channel fading models with various adaptive transmission policies under different fading and transmission constraints. Below is the description of different generalized fading conditions with EC metric analysis in different wireless networks:   }


Under the generalized fading model, $\alpha-\mu$ fading model uses   $\alpha-\mu$ distribution. As compared to  stochastic fading models, $\alpha-\mu$ distribution provides more generality to analyse the fading environment  \cite{R17}. In  \cite{R20}, authors have used  \blue{the  EC model}   in underlay CRNs\footnote{for details on underlay CRNs, see Section \ref{sec:networks}.}  with $\alpha-\mu$ fading model. In this study, an in-depth performance analysis based on symbol error probability using EC concept has been  performed with $\alpha-\mu$ fading conditions.  \blue{In this work, the EC model has been employed to understand the peak interference power to noise ratio. The results suggest that  when the delay becomes more stringent, the peak interference power to noise ratio becomes smaller. However, the peak interference power to noise ratio increases with the increase in  EC. This analysis is useful for understanding the performance of cognitive cooperative  networks over $\alpha-\mu$ fading channels.  }


How  shadowing  \blue{affects} the performance in wireless networks, can also be investigated with the help of $\kappa-\mu$ fading conditions. \blue{$\kappa-\mu$ fading model provides \blue{a} more  general model by covering  Gamma shadowing, one-sided Guassian, Nakagami-m, Rayleigh, and Rician  fading.} Authors in  \cite{R24,R195}, have discussed   $\kappa-\mu$ fading model and analyzed the system gain  through the concept of EC.   Analytical expressions for MISO systems with \blue{the  EC model} under $\kappa-\mu$ fading conditions have been provided.  \blue{MGF approach is employed to investigate the achievable EC of MISO systems. This work provides a useful insight into the MISO systems, by investigating the asymptotic analysis of the EC at high SNR. Without the concept of EC, these insights are not possible with any other model. } 

\amj{
EC-based QoS analysis has also been performed with $\kappa$ fading channels to test the performance of   shadowing and multipath propagations and their impact on  received signal \blue{quality}.  Impact of generalized $\kappa$ fading conditions in cooperative communications has been analyzed with EC concept in \cite{R56}. Under the influence of generalized $\kappa$ fading conditions, it has been estimated that,  increasing the number of relays can also maximize the performance of a system with  delay constraints. }

\subsection{Futuristic Models}
Some futuristic models such as  \amj{matrix-exponential distribution (MED)}    and broadcast fading models have also been investigated with EC   \amj{model} in more detail. With the help of futuristic models, some aspects of wireless communications such as  retransmission and compression schemes have been   investigated with \blue{the  EC model}.   Below is the description of some versions of futuristic fading models \amj{that have been investigated with EC}. 
\subsubsection{MED} 
Different versions of retransmission schemes such as persistent and truncated ARQ and HARQ have been investigated with  MED-based  channel model \blue{in} \cite{R06}. An EC-based analytical expression has been provided for different retransmission schemes\footnote{for details on retransmission scheme, see Section \ref{sec:retransmission}.}  while considering MED-based  fading channel. With the help of this fading condition and \blue{the  EC model},  impact of diversity (due to MIMO antennas) on the system has been   \amj{investigated}  in detail. \blue{This work also provides the EC expression for the persistent truncated, and networked HARQ, that was not addressed before. Under the MED-based fading channel and using the EC model, the target delay with delay violation probability has been estimated. This can lead to obtaining the important results regarding the general effective channel functions and transition probability. Another important take-away from this study is that, the diversity of the MIMO under MED fading channels, reduces the sensitivity of EC with respect to different delay exponents.}

\subsubsection{Broadcast Fading Models}
 \amj{Fading in  broadcast or multicast channels  can also be investigated with the help of \blue{the  EC model} \cite{R252}}. The EC  \amj{concept}  with broadcast fading channels in mobile  networks has been  \amj{investigated}  in \cite{R96}. In broadcast channel,  different fading states across the receiver has been taken into account while  estimating the performance of a system with EC. A trade-off has been developed among delay constraints, QoS guarantee, throughput, and reliability. In this study,  reliability has  also been taken into account  \amj{to estimate the overall packet loss.} 

\subsubsection{Model for Unknown Fading Conditions}
 \amj{In some cases, there are scenarios, where  statistics of a fading is unknown.} \amj{For}  such links where  fading statistics are unknown, authors in \cite{R155} have proposed a scheme where  channels with unknown statistics can be modelled using  \blue{the  EC model}.   This scheme \amj{also}   takes into consideration   \amj{an} optimal power allocation and link selection \amj{scheme} with EC.     This introduces more flexibility in  \amj{a} system, as proposed approach has the capability to converge in any fading distributions depending upon the  transmission scenarios and fading.    

\subsection{Dimension-Based Models}
Dimension-based models are also called  multi-ray fading models. As compared to other fading models, these models calculate  path-loss  along all possible  paths or depending upon diverse fading conditions.  The existing work  considering  dimension-based fading models with  \blue{the  EC model}  can be classified into two-dimensional (2D) and multi-dimensional \amj{fading models}.    
\subsubsection{2D Fading Model}
  Achievable EC of wireless networks    \blue{under}    2D fading channel has been  \amj{investigated}  in \cite{R91}. In this study, \blue{a} 2D-based  Markov model has been used to investigate the fluctuating  channel \amj{conditions}   QoS at link-layer has been  analyzed \amj{with EC}. \amj{For this purpose,}  arrival rate\amj{/source rate is find with EC and when the  \blue{arrival} rate is fixed, the delay experienced by the arriving packet is estimated. }   Proposed scheme has been tested extensively with \blue{the  EC model} and compared with other schemes.  

\subsubsection{Multidimensional Fading Models}
As compared to \cite{R91}, in which  2D-based Markove process has been used to model the fading conditions,  in  \cite{R95} authors have employed  \blue{a} multi-dimensional-based  Markov process to  model  fading conditions and then used the EC model to investigate the delay \blue{outage probability}.   In this study, a cross-layer \amj{resource management approach has been used with \blue{the  EC model} to investigate}  arrival-rate, queueing behavior,  \blue{in}   \amj{multidimensional fading channels.}   \amj{Extensive simulation shows that} proposed scheme   \blue{achieves higher}  throughput while  guaranteeing the required delay for QoS-aware applications.   

\subsection{Supplementary Models}
Supplementary fading models have also been analysed with \blue{the  EC model}. Supplementary models are not the  distinct class of fading \amj{models,}   but they are usually introduced to address  certain   \amj{limitations} of existing fading channels.  Supplementary fading models  are actually based on  existing fading models. Below is the description of the some supplementary fading models \amj{with EC theory}, that has been proposed to address  some special features or aspects in existing fading channels/models:
\subsubsection{Turbulence Model}
In   \blue{advanced} wireless communications \amj{systems},  such as optical wireless communications (OWC),  impact of turbulence fading  has also been investigated with \blue{the  EC model} \cite{R07}.  \amj{Following works \cite{R07,R131,R147,R148} consider the turbulence fading  with their achievable EC in OWC .} In these studies   turbulence fading conditions have only been analyzed for the OWC \amj{with different power-adaptation schemes. The closed-form expression of the EC with turbulence fading has also been derived and validated through extensive simulations.}   

\subsubsection{Gamma Fading}
 As compared to turbulence fading, Gamma-Gamma turbulence fading has also been discussed in OWC in  \cite{R16}. With the help of this fading distributions, the independent and joint power adaptation in OWC with their achievable EC   \blue{have}  been  \amj{investigated}  in detail. From this study, it has been clear that if the fading is  \amj{minimized} and  \amj{a} delay constraint is loose, then the performance gap between  independent and joint power adaptation is minimal. Authors in \cite{R54}, have utilized  Gamma distribution for modelling  wireless channel in CRNs with EC.   Optimal power allocation with power/interference-power has been analyzed with statistical QoS provisioning.  Proposed scheme has been extensively tested with  \amj{its achievable EC through simulations}  and shows  improved performance as compared to  other \amj{state-of-the-art} DSA techniques. 

\subsubsection{Temporal Fading}
Supplementary \amj{fading} models  with temporal distributions  \blue{have}   been  \amj{investigated}  in smart grid environment  \cite{R19}. In this study,  \amj{the performance}  of wireless communications systems  in smart grid \amj{environment} has been  \amj{investigated with the EC model.}   Fading conditions have been modeled using \blue{a}  temporal fading model. This EC-based delay analysis of  smart grid's communications architecture shows  adaptability of this architecture for different smart grid's applications.

\subsection{Summary and Insights}
\label{subsec:fading_summary}
\amj{Complexity of  \blue{the  EC model} increases as the fading conditions become more and more complex. As compared to the ergodic and Shannon capacity, the closed form expression for EC by taking into consideration the different fading conditions is relatively more complex and difficult \blue{to obtain}. This is a serious challenge for considering the EC over other capacities for testing the performance of any systems. EC has been extensively used with the i.i.d Rayleigh fading conditions, however, the accuracy of the EC model with non i.i.d fading conditions invites the future researchers to investigate further. }

 \blue{The} EC \blue{model} helps in understanding QoS-requirements with varying service rate under different fading conditions.  Changes in EC according to different fading conditions  \blue{provides a strong mathemtiacal tool for observing the effects of fading conditions on the delay performance of the considered fading channel.}   Through these changes, optimal resource  \blue{allocation} schemes, scheduling algorithms, and network designs can also be \blue{proposed and} investigated in more detail.

\section{Effective capacity Measures in Different Networks}
\label{sec:networks}
With the help of \blue{the  EC model}, \blue{the performance of}  different wireless networks  under different delay requirements  \blue{can be  investigated}  by considering  \amj{various} wireless designs and architectures.  \amj{Achievable EC of}  CRNs, wireless sensor networks,  relay-networks, and mesh networks have been analysed extensively in the   \blue{literature}.   In this section, we have broadly discussed these networks with their different design and architectural aspects with the EC \amj{model.} 

\subsection{Cognitive Radio Networks  (CRNs)}
\label{subsec:crns}

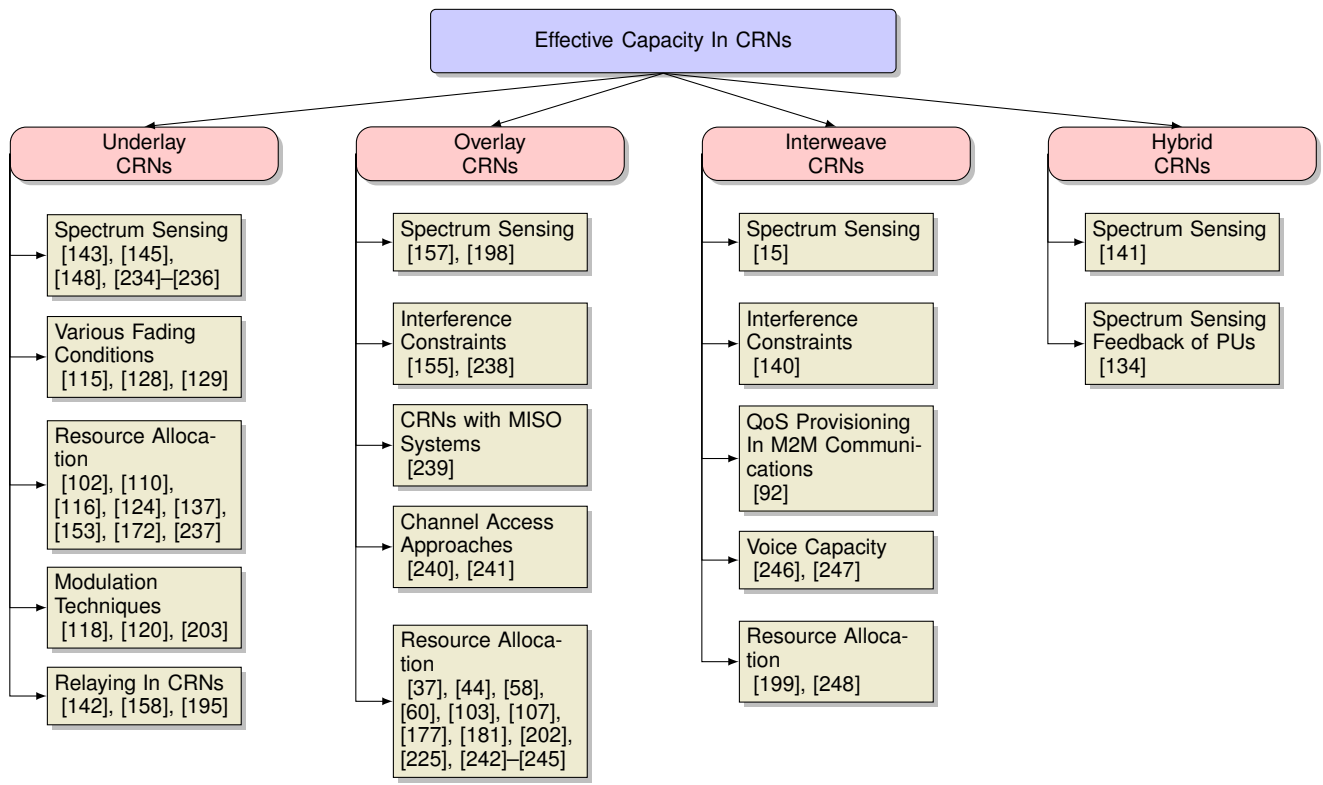
\begin {figure*}
\footnotesize
\centering
\begin{tikzpicture}[
  level 1/.style={sibling distance=46mm},
  edge from parent/.style={->,draw},
  >=latex]

\node[root,minimum height=3em,text width=6cm,fill=blue!20] {Effective Capacity  In CRNs}
  child {node[level 2] (c1) {Underlay \\CRNs}}
  child {node[level 2] (c2) {Overlay \\CRNs   }}
  child {node[level 2] (c3) {Interweave\\CRNs }}
  child {node[level 2] (c4) {Hybrid \\ CRNs }};
\begin{scope}[every node/.style={level 3}]
\node [below of = c1, xshift=0pt,yshift=-10pt] (c11) {Spectrum Sensing \\ \cite{R04,R34,R42,R55,R168,R214}  };
\node [below of = c11,yshift=-10pt] (c12) {Various Fading Conditions \\ \cite{R17,R20,R202}};
\node [below of = c12,yshift=-20pt] (c13) {Resource Allocation \\\cite{R23,R59,R76,R82,R150,R162,R303,R216} };
\node [below of = c13,yshift=-18pt] (c14) {Modulation Techniques \\\cite{R236,R253,R258} };
\node [below of = c14,yshift=-5pt] (c15) {Relaying In CRNs \\\cite{R32,R87,R213} };
\node [below of = c2,xshift=0pt,yshift=-5pt]	 (c21) {Spectrum Sensing \\\cite{R85,R225}
};
\node [below of = c21,xshift=0pt,yshift=-10pt] (c22) {Interference Constraints \\\cite{R83,R304}};
\node [below of = c22,,xshift=0pt,yshift=-10pt] (c23) { CRNs with MISO Systems\\ \cite{R232}};
\node [below of = c23,,xshift=-0pt,yshift=-10pt] (c24) {  Channel Access Approaches \\ \cite{R181,R197}}; 
\node [below of = c24, xshift=0pt,yshift=-30pt] (c25) {Resource Allocation\\\cite{R54,R88,R92,R122,R124,R134,R142,R163,R169,R171,R180,R229,R243,R256}};
\node [below of = c3, xshift=00pt,yshift=-5pt] (c31) { Spectrum Sensing \\\cite{R03} };
\node [below of = c31, yshift=-10pt] (c32) {Interference Constraints \\ \cite{R28} };
\node [below of = c32, yshift=-15pt] (c33) {QoS Provisioning In M2M Communications  \\ \cite{R159} };
\node [below of = c33, yshift=-10pt] (c34) {Voice Capacity  \\ \cite{R227,R241} };
\node [below of = c34, yshift=-10pt] (c35) { Resource Allocation\\ \cite{R224,R245} };
\node [below of = c4, xshift=0pt,yshift=-5pt] (c41) {Spectrum Sensing  \\ \cite{R31} };
\node [below of = c41,yshift=-10pt] (c42) {Spectrum Sensing Feedback of PUs\\ \cite{R09}};

\end{scope}

\foreach \value in {1,2,3,4,5}
  \draw[->] (c1.180) |- (c1\value.west);

\foreach \value in {1,...,5}
  \draw[->] (c2.180) |- (c2\value.west);

\foreach \value in {1,...,5}
   \draw[->] (c3.180) |- (c3\value.west);
   
\foreach \value in {1,2}
  \draw[->] (c4.180) |- (c4\value.west);
%
   
\end{tikzpicture}
\caption {Effective capacity in CRNs can be studied while employing various white space paradigm such as overlay, underlay, Interweave, and hybrid CRNs.   }
\label{fig:crns-ec}
\end{figure*}

Current spectrum crunch can be avoided with the help of cognitive radio networks (CRNs). CRNs employ  dynamic spectrum access (DSA) approach and dynamically assign  spectrum resources while avoiding the   \amj{interference} to  primary users \cite{R197}.  Cognitive users that use  licensed spectrum band are termed as the secondary users (SUs), while  users of licensed spectrum band are named as the primary users (PUs). In CRNs, SUs utilize the idle portion of licensed spectrum band while keeping  minimum interference to PUs.  Idle portions of    licensed band are  \amj{termed}  as white spaces \cite{Saleem14jnca,R317}.

Various QoS-aware applications in CRNs can well be tested with the help of EC metric  \cite{R162}.   \amj{Source and service rate} characterization is an important feature of the EC model. This feature has been well-\amj{exploited} in CRNs.  With  achievable EC in CRNs, various  \amj{techniques} such as efficient spectrum sensing, resource allocation, modulation, spectrum access, and interference constraints can be analyzed.

Achievable  EC  in CRNs has been studied from the perspective of   white space \blue{utilization}. According to the utilization of white space,  CRNs can be categorized into underlay, overlay, interweave, and hybrid CRNs~\cite{Akhtar16tp}, \amj{\cite{R150,R54}}. Figure \ref{fig:crns-ec} shows the classification of existing work on EC-based CRNs into underlay, overlay, interweave, and hybrid CRNs.

Table \ref{tab:crns-ec} shows  various gains with their corresponding study in different categories of  CRNs with the EC model.  Impact of PUs activity patterns  in different classes of CRNs has also been \amj{presented}. Below is the description of different classes of CRNs   with the concept of EC.

\begin{table*}
  \footnotesize
\centering
\caption{Effective Capacity-Based Performance analysis In Cognitive Radio Networks (CRNs) }
\label{tab:crns-ec}
\begin{tabular}{|p{2cm}|p{0.7cm}|p{0.8cm}|p{4.5cm}|p{2cm}|p{4.5cm}|}
\hline
\bfseries White Space Paradigm  & \bfseries Study & \bfseries PUs Activity  & \bfseries  Factors evaluated with EC metric& \bfseries Fading Channel Model & \bfseries Area   \\
\hline
 
Underlay CRNs &\cite{R34}     & & Channel  Errors   &Rayleigh& Spectrum Sharing with Imperfect CSI\\\cline{2-6}
 &\cite{R42}     &\checkmark &QoS  with Interference Limitations   &Rayleigh&Multiple Channel Spectrum Sharing \\\cline{2-6}
 &\cite{R55}     &\checkmark &Interference Limitations     &Rayleigh&Spectrum Sensing In CRNs\\\cline{2-5}
   &\cite{R168}     &\checkmark & Channel Gain     &Rayleigh&\\\cline{2-5}
   & \cite{R214}  & &Analysis of Outdated CSI    &Not Defined& \\\cline{2-6}
  &  \cite{R17}   & &  General Fading    &$\alpha-\mu$&CRNs with Various Fading Conditions  \\\cline{2-5}
 &\cite{R20} & &  General Fading     &$\alpha-\mu$& \\\cline{2-5}
 &\cite{R202}     & &Different QoS Applications    &Nakagami-$m$&\\\cline{2-6} 
  &\cite{R04}     &\checkmark &Optimal Power Allocation    &Rayleigh&Resource Allocation  \\\cline{2-5}
 &  \cite{R23}  & &Trade-off Between   SUs and PUs performance  & Rayleigh& \\\cline{2-5}
  &  \cite{R59}    & \checkmark&Interference Power Constraint     &Gamma Fading& \\\cline{2-5}
 &             \cite{R76}    & &Power Allocation with Imperfect CSI   &Rayleigh& \\\cline{2-5}
&             \cite{R82}    & \checkmark& Power Allocation with Imperfect CSI   &Rayleigh& \\\cline{2-5}
   &     \cite{R150}     & &Securing the CRNs   &Nakagami-$m$&\\\cline{2-5}
               &  \cite{R162}     & &Optimal Power Allocation   &Rayleigh& \\\cline{2-5} 
                              & \cite{R216}   &\checkmark &Optimal Power Allocation   &Nakagami-$m$&\\\cline{2-6}
                 & \cite{R303}  & &Optimal Power Allocation  &Nakagami-$M$&CR-Based MAC  \\\cline{2-6}
            &       \cite{R236}     &\checkmark &Rate and Power Estimation with MQAM    &Nakagami-$m$&Modulation In CRNs\\\cline{2-5}
                                  &   \cite{R253}    &\checkmark & Power Estimation with MQAM    &Nakagami-$m$&\\\cline{2-6}

    &      \cite{R258}    & & SUs Throughput    &Rayleigh&FDMA-Based CRNs\\\cline{2-6}
  &    \cite{R32}    & & SUs arrival rate   &Rayleigh&CR-Relay Networks \\\cline{2-5}
&     \cite{R87}&\checkmark &Interference and Delay Constraints    &Rayleigh&\\\cline{2-5}
 &    \cite{R213}& & SUs Capacity    &Rayleigh& \\\hline   
                  
Overlay CRNs&  \cite{R85}     &\checkmark &Power Level based on PUs    &Rayleigh& Spectrum Sensing In CRNs\\\cline{2-5} 
& \cite{R225}  &\checkmark & SUs Arrival Rate    &Rayleigh&  \\\cline{2-6}
&     \cite{R83}  & &Performance Limitations of SUs    &Rayleigh&Interference Limitations In CRNs\\\cline{2-5}
 &   \cite{R304}     &\checkmark &SUs Capacity Analysis    &Asymmetric Fading&\\\cline{2-6}
 &    \cite{R232}     & &Channel Estimation Error   &Nakagami-$m$& CRNs with MISO Systems\\\cline{2-6}
 &    \cite{R181}    &\checkmark &Two phase channel Access Method    &Not Defined&Channel Access In CRNs\\\cline{2-5}
&    \cite{R197}  &\checkmark & Hopping Based Channel Access    &Not Defined&\\\cline{2-6}
 &    \cite{R54}    &\checkmark &Optimal Power Allocation  &Gamma Fading&  Resource Allocation\\\cline{2-5}
 &    \cite{R88}  & & SUs Arrival Rate    &Nakagami-$m$&\\\cline{2-5} 
 &\cite{R92}   &\checkmark &Optimal Power Allocation    &Nakagami$m$&\\\cline{2-5}
&    \cite{R122} & &Optimal Resource Allocation    &Rayleigh&\\\cline{2-5} 
 &    \cite{R124}  &\checkmark &Optimal Power Allocation    &Nakagami-$m$&\\\cline{2-5}
 &    \cite{R134}  &\checkmark & Full-Duplex Operation     &Nakagami-$m$&\\\cline{2-5} 
&\cite{R142}    & &Optimal Power Allocation   &Rayleigh& \\\cline{2-5}
 &\cite{R163}    & &Power Consumption    &Rayleigh&\\\cline{2-5}           
 &\cite{R169}    &\checkmark &Optimal Resource Allocation    &Not Defined&\\\cline{2-5}   
 &\cite{R171}    &\checkmark &Optimal Time Slot Allocation    & Rayleigh&\\\cline{2-5} 
 &\cite{R180}    & &Optimal Channel and Power Allocation    &Rayleigh&\\\cline{2-5}
 &\cite{R229}    & &Cross-tier Interference    & Rayleigh&\\\cline{2-5}
 &\cite{R243}    & &Cross-tier Interference    &Rayleigh&\\\cline{2-5} 
 &\cite{R256}    & &QoS-Based Power Allocation    &Rayleigh&\\\hline 
 
Interweave CRNs	& \cite{R03} &\checkmark&Optimal Sensing Time Achieved&Nakagami-$m$&Spectrum Sensing In CRNs  \\\cline{2-6}
 
                   & \cite{R28}  &\checkmark &Interference Power Limitations    &Rayleigh&Interference Constraints\\\cline{2-6}
                    
               & \cite{R159}     &\checkmark &QoS in Cognitive M2M  &Rayleigh&Cognitive M2M Communications  \\\cline{2-6}
              &    \cite{R227}  &\checkmark & Voice Capacity    &Not Defined&Voice Capacity in CRNs\\\cline{2-6}
               & \cite{R241}     &\checkmark & Voice Capacity    &Rayleigh&Wireless Multimedia CRNs\\\cline{2-6}
                    &\cite{R224} &\checkmark &Delay Bound  &Not Defined& Resource Allocation \\\cline{2-5}
                       & \cite{R245}     &\checkmark&SUs Capacity  &Rayleigh&  \\\cline{2-6}                
              &    \cite{R41}     &\checkmark &Optimal SUs Transmission Rate    &Nakagami-$m$&CRNs performance Without CSI\\\cline{2-6}
 
         &       \cite{R70}    & &SUs performance with Imperfect CSI    &Rayleigh&GSC In CRNs\\\hline 
                  Hybrid CRNs &  \cite{R09}  &\checkmark &Complexity and Cost   &Rayleigh&PUs Spectrum Sensing Feedback \\\cline{2-6} 
 &\cite{R31}    &\checkmark &SUs performance Quantified    &Rayleigh&Spectrum Sensing In CRNs\\\hline

                                                           \end{tabular}
\end{table*}

\subsubsection{Underlay CRNs}

This white space utilization pattern is also named as  gray-space utilization \amj{in CRNs}. In this  \amj{scheme,}   SUs  can transmit simultaneously with PUs, while keeping the interference to  PUs within \blue{an} acceptable range. For this purpose, SUs  \amj{have to use}  low-power cognitive devices with much limited range compared to other classes of CRNs.  Spectrum sensing, modulation, and  optimal resource  \blue{allocation} under different fading conditions in underlay CRNs  \amj{have been well-investigated with the EC model.}  \amj{Underlay CRNs with their achievable EC with different spectrum sensing approaches have  been discussed in \cite{R04,R34,R42,R55,R168,R214}.}  Most of the work on spectrum sensing with \blue{the  EC model}  considers  energy-detection-based spectrum sensing \amj{approach}  with predictable and known parameters. However, other spectrum sensing approaches, such \amj{as} QoS-aware, cooperative, and cyclostationary-based spectrum sensing have not been explored while taking into consideration the EC model. 

This gray-space utilization pattern of CRNs with \blue{the} EC concept has also been used to  \amj{develop various}  optimal resource allocation schemes \amj{with stringent-delay requirements} \cite{R23,R59,R76,R82,R150,R162,R303,R216}. Among  resource allocation strategy,  problem of power allocation has been formulated and then solved by applying different optimization scenarios. However,  radio resource allocation schemes \amj{(spectrum assignments)}  with the  EC model for underlay CRNs have not been studied \amj{in detailed}.

Statistical QoS-provisioning in underlay CRNs with \blue{the} EC concept  has also been studied in CR-based \amj{cooperative communications}  with single and multiple relays \cite{R32,R87,R213}. In these work,  buffer-aided relaying strategy has  been employed.   \blue{The  EC model}   in these  underlay CRNs has also been used with varying fading conditions,  such as generalized fading models,  stochastic \amj{fading} model,   \blue{and}  Nakagami-$m$ fading channel \cite{R17,R20,R202}.   \blue{Adaptive}  modulation techniques have also been analyzed  \blue{using the} EC concept in underlay CRNs  \cite{R236,R253,R258}. In these modulation schemes,  average power consumption has been monitored with the provision of statistical QoS.  Average and available rate have also been analysed while keeping the interference to  PUs within minimum range. 

\subsubsection{Overlay CRNs}
In overlay CRNs,  SUs can exploit  licensed spectral resources either cooperatively or non-cooperatively, with or without the presence of PUs. SUs can simultaneously transmit with PUs by adjusting their transmission parameters to keep the  \amj{interference}  to PUs at a minimal and  acceptable limit. As compared to  underlay CRNs, in overlay CRNs,  PUs can  acknowledge the  \amj{presenece} of SUs at   \amj{a} licensed spectrum band (in case of cooperative communications) \cite{R317}. 

The EC \amj{model} with overlay CRNs has been used while analysing different aspects \amj{of overlay CRNs,} such as spectrum sensing, resource allocation, and channel accessing methods. EC concept with overlay CRNs has been used with different spectrum sensing approaches such as discussed in \cite{R85,R181,R197,R225}. In these work, efficient spectrum sensing approaches based on required delay-outage probability has been investigated with the achievable EC while keeping the interference to PUs at \blue{a} minimal range \blue{are obtained}.  Optimal resource  \blue{allocation} schemes in  overlay CRNs \cite{R54,R88,R92,R122,R124,R134,R142,R163,R169,R171,R180,R229,R243,R256} with interference constraints \cite{R83,R304} has also been \amj{investigated} with EC.   In addition to these \amj{resource allocation schemes}, FD communications behavior for overlay CRNs have also been analyzed  with \blue{the  EC model}. Antenna diversity  with multiple input and multiple output (MISO) antennas   in  overlay CRNs has been analysed with \blue{the} EC concept  \cite{R232}. In this study,  channel estimation errors have been   evaluated with their impact on the performance of overlay CRNs with MISO antennas.

\subsubsection{Interweave CRNs}
When there is no PUs activity  \blue{in}    a licensed band,  white space utilization \amj{by SUs in this case}  is termed as  \amj{an} interweave  \amj{white space utilization or interweave CRNs}  \cite{R317,R70}. In interweave CRNs, SUs can only utilize  licensed spectrum resources when  PUs is idle.

Most of the heterogeneous real-time applications are supported by  interweave CRNs. EC concept in interweave CRNs has also been used to analyse the statistical QoS provisioning with delay constraints for different real-time, QoS-aware, and bandwidth hungry applications such \blue{as}  multimedia applications  \cite{R227,R241,R159}. Energy-detection based spectrum sensing \cite{R03}, interference constraints \cite{R28}, and resource allocation \amj{scheme} \cite{R224,R245},  have also been analysed with \amj{achievable} EC  in interweave CRNs.  Impact of PUs activity has also been \amj{studied}  with \blue{the} EC concept. However, different PUs activity patterns, such as low, high, and intermittent PUs activity with their impact on interweave CRNs  has not been \amj{investigated } while considering \blue{the  EC model}. This opens the new vistas for  future research directions.

\subsubsection{Hybrid CRNs}
In most  cases, SUs can exploit  idle spectrum resources by adopting  \amj{a}  hybrid approach of white space utilization (combination of above two or three) \cite{R317}. By applying  \amj{a}  hybrid approach,  limitations of above mentioned approaches can be  \blue{overcomed}. However, this approach introduces some type of  complexity in managing different  \amj{factors such as PUs activity and spectrum mobility}. Authors in \cite{R31,R09}   \amj{have modelled the queueing-delay with  the help of EC }   in hybrid CRNs with spectrum sensing paradigm. In this study,  feedback from  PUs for interference has also been analyzed. This feedback is then taken into account for  designing  \amj{an}  optimal scheduling approach.  Queue length information with PUs feedback is used for  \amj{guaranteeing}  the statistical QoS with efficient spectrum sensing approach. \blue{A tradeoff between the SUs achievable EC and the PUs success rate has been characterized. From this tradeoff, a three level power allocation scheme is then developed to enhance the QoS-aware performance of the SUs.}

\subsection{Cooperative Networks}
Due to   \amj{ the advancement} in relaying protocols, \amj{cooperative networks}   have gained much attention.  \amj{In cooperative networks,}   \amj{a}  source node can communicate with  \amj{a}  destination node with some intermediate nodes/relays.   \amj{Statistical} QoS-provisioning has been studied by characterizing the traffic not only on  \amj{a}  source node but also at  \amj{a} relay node  \cite{R208}. Table \ref{tab:relay-ec} shows the recent work based on \blue{the  EC model} in  \amj{cooperative}   networks.  Existing work on EC-based QoS estimation in  \amj{cooperative}  networks has been classified based on the relay selection approaches. 

In traditional  \amj{cooperative} networks,   relaying is performed either in regenerative or non-regenerative way. When  \amj{a}  regenerative approach is considered,  \amj{a} decode-and-forward (DF) relaying method is used, while non-regenerative approach employs the amplify-and-forward (AF) scheme in selecting  \amj{a}  relay node.   Advance wireless networks also consider other relaying approach such as QoS-aware relaying, buffer-aided relaying, and detect-and-forward relaying approach.  \amj{Achievable EC of cooperative networks with different relaying schemes has been surveyed below:}

\begin{table*}
  \footnotesize
\centering
\caption{Effective Capacity-Based Performance Testing In Cooperative Networks}
\label{tab:relay-ec}
\begin{tabular}{|p{4cm}|p{4cm}|p{1cm}|p{6cm}|}
\hline
\bfseries Relaying Method& \bfseries Architecture or Design Involved    & \bfseries Study  & \bfseries  Factors evaluated with EC metric\\
\hline

Amplify-and-Forward 	 &Multi-User Cooperative Network& \cite{R05}&Optimal Resource Allocation\\\cline{2-4}
 
   	&Multi-Hop Network Design& \cite{R12}&EC performance Analysis with Channel State Information\\\cline{2-4}
     	&Multi-Relay Networks&  \cite{R40} &Optimal Resource Allocation\\\cline{2-4}
   
    	&Two-Way Relay Networks& \cite{R44} &Analysis of Various Gains of Two-Way Relay Networks\\\cline{2-4}
 
   	&Full-Duplex OWC Network&    \cite{R131}&Optimal Resource Allocation with Throughput Maximization\\\cline{2-4}  
   	&Single Relay Network&    \cite{R149}&Cross-Layer Power Allocation\\\cline{2-4}
   	&Virtualized Relay Networks&    \cite{R161}&Optimal Power Allocation\\\hline

Decode-And-Forward    	&Multi-Relay Networks& \cite{R48} & Relaying Scheme with Modulation\\\cline{3-4}
   	&&     \cite{R65} &EC Analysis With Four Retransmissions Schemes\\\cline{2-4}
     	&Full-Duplex Relay Networks&  \cite{R133}&Optima Resource Allocation\\\cline{2-4}    
  
&Three-Mode Relay Networks& \cite{R62}    	& Adaptive Relaying\\\hline 

Buffer-Aided    	&Three Node Relay Networks&\cite{R15} &EC-Analysis with Buffer Aided Relaying\\\cline{3-4}
  	&&\cite{R166} &EC-Analysis with Buffer Aided Relaying\\\cline{2-4}
     	&Full-Duplex Relay Networks&\cite{R21}&EC-Analysis with Buffer Aided Relaying\\\cline{2-4}
  	&Two-Hop Networks&\cite{R30}&Resource Allocation\\\cline{3-4}
   	& &\cite{R36}&Concurrent Relay Selection Has been proposed\\\cline{3-4}
   	  	&&\cite{R72} &Buffer Constraints and Throughput Has been explored\\\cline{2-4}
   	&Diamond Relay Networks with two Relays&\cite{R127}&Performance analysis with New Selection Policy\\\cline{3-4}
	&&\cite{R130} &Performance analysis with New Selection Policy\\\hline

 QoS-Based    	&Multi-Relay Networks&   \cite{R94}&Optimal Resource Allocation\\\cline{2-4}
    	&Heterogeneous Relay Networks&    \cite{R200}&Optimal Resource Allocation\\\cline{2-4}
    	&Two-way Relay Network&  \cite{R215} &Optimal Cross-Layer Resource Allocation\\\cline{2-4}
  &Mobile Multi-Hop Relay Networks&  \cite{R231}&Cross-Layer Simulation Platform  \\\cline{2-4}
  &Two-Hop Networks &  \cite{R272}&TDMA and OFDMA-Based Resource Allocation \\\cline{3-4}
 && \cite{R273} &TDMA and OFDMA-Based Resource Allocation \\\hline 

 Frequency-Based     	&OFDMA-Based Relay Networks& \cite{R210}&Optimal Resource Allocation\\\cline{3-4}
 	&&   \cite{R217}  &Optimal Resource Allocation\\\cline{2-4} 

  & CR-Based Relay Networks&\cite{R225}&Cross-Layer Resource Allocation with Imperfect Spectrum Sensing \\\hline

Amplify-And-Forward, Decode-And-Forward  &Full-Duplex Relay Networks&\cite{R221}&Optimal Resource Allocation \\\cline{2-4}
  &Two-Hop Relay Networks&\cite{R274}&Cross-Layer Resource Allocation \\\hline
Detect-And-Forward  	&Multi-Relay Networks&       \cite{R99} &Optimal Resource Allocation\\\hline 

                                                            \end{tabular}
\end{table*}

\subsubsection{Amplify-and-Forward Relaying}
Achievable  EC  for non-regenerative approach of relaying named as amplify-and-forward (AF) relaying scheme has  been discussed in  \cite{R144}. EC \amj{of}   AF relaying scheme  \amj{has been studied extensively in  existing work}  as compared to other relaying scheme such as DF  \cite{R144}.  DF relaying performs some type of processing before sending the packets.   The concept of achievable EC in AF relaying has also been exploited to evaluate the performance of  FD cooperative communications  \cite{R131}, multi-user cooperative networks \cite{R05}, multi-hop networks \cite{R12,R40}, two-way relaying \cite{R44}, single-relaying \cite{R149}, and virtualized relaying \cite{R161}.   AF relaying protocol with EC concept in most cases considers  optimal power allocation schemes to  \amj{achieve} energy conservation  in energy-scarce networks. However, other issues such as radio resource allocation in  CR-based relay networks has not been investigated with \blue{the  EC model}. \amj{This invites the future researchers to investigate further in this domain.}

\subsubsection{Decode-and-Forward Relaying}
Decode-and-forward (DF) relaying is  \amj{a}  regenerative relaying approach that regenerates or decode the packets \amj{first} before transmission to another node in  \amj{cooperative}  networks.  \amj{Investigating the achievable EC of}  DF relaying scheme is   more complex as compared to  AF relaying.  DF relaying is quite similar to  ``processing base station'', hence  characterization of \amj{source} traffic   based on delay constraints on  processing base  \amj{station} or relay node adds more complexity in the system.   This complexity has been minimized by adopting   \amj{an} advance DF-base relaying network architectures such  as FD-relaying \cite{R133}, three-mode relaying \cite{R62}, and multi-relaying \cite{R48,R65}.

\subsubsection{Buffer-Aided Relaying}
Buffer-aided relaying is  \amj{a} sub-type of AF and DF relaying. Buffer-aided relaying can be performed by adopting any relaying protocol (AF or DF) with additional buffers \cite{R15,R155}.  \amj{Cooperative}  networks with buffer-aided relaying can support variable \amj{source} rate by adjustable buffers at  source and the relay nodes and their performance has been tested with \blue{the  EC model}.  EC concept in buffer-aided relaying  with advance communications designs/architectures such as three-node relay networks \cite{R15,R166}, FD-relay networks \cite{R21}, two-hop networks \cite{R30,R36,R72}, and diamond-relay networks \cite{R127,R130} has also been used to  \amj{investigate}  the performance of these advance networks with \blue{under} delay-violation probability \blue{constraints}.     

\subsubsection{QoS-Based Relaying}
QoS-based relaying can  be performed with the help of regenerative, i.e., DF or non-regenerative, i.e., AF approach of relaying. This relaying approach has specially been introduced to incorporate the demand of  \amj{stringent delay requirements}  at the relay nodes.  EC concept has also been  \amj{investigated}  with this relaying approach by ensuring  statistical QoS  provisioning. This relaying approach with EC concept  has  further been used to analyze   multi-relays \cite{R94}, heterogeneous relays \cite{R200}, two-way relaying \cite{R215}, mobile multi-hop relaying \cite{R231}, and two-hop networks \cite{R272,R273}. QoS-based relaying shows improved performance as compared to  other relaying schemes when considering  multimedia applications with stringent delay requirements. \blue{These studies provide  important insights, such as, the EC can be improved by properly adjusting the location of the relays. Furthermore, when the relay networks are investigated with the FDMA and TDMA approaches with their achievable EC, it has turned out that  FDMA approache outperforms   TDMA. }

\subsubsection{Frequency-Based Relaying}
Relaying can also be performed by considering  frequency domain of  \amj{a} wireless communication paradigm. By considering  frequency as  \amj{a}  decision factor for selecting the relay,  issue of spectrum scarcity can also be addressed \cite{R210,R107}.  Multimedia applications with the help of \blue{the  EC model} has also been analyzed in frequency-aware relaying. OFDMA-based relay networks \cite{R210,R217} and CR-relay networks \cite{R225} consider  frequency-based relay selection \amj{approach} and employ \blue{the  EC model} to test the performance of   \amj{a system}. 

\subsubsection{Amplify-and-Forward and Decode-and-Forward Relaying}
Hybrid relaying approach comprising of AF and DF relaying approach  has also been used in many advance  wireless networks. With   \amj{a} hybrid approach,  limitations of AF and DF relaying schemes can be addressed while supporting  QoS-aware applications with  specific delay  requirements \cite{R274}. This approach of relaying in conjunction with FD-relaying \cite{R221} and two-hop relaying \cite{R274}  has also been analysed with EC concept \amj{with the required delay-violation probability}.  \blue{Achievable EC of both relaying schemes has been compared with the direct transmission while taking into consideration the stringent delay requirements. This analysis shows the superior performance of  both relaying schemes as compared to the direct transmission when  delay requirements are  stringent.}

\subsubsection{Detect-and Forward Relaying}
In comparison to  AF and DF relaying, detect-and-forward (DeF)   relaying has also been \amj{analyzed} with the EC \amj{model}. In \cite{R99}, DeF relaying approach takes into consideration the achievable EC in multi-relay networks. First, the best channel is detected  \amj{on that basis}  the relay is selected. With the increase in  QoS-requirements \amj{with stringent delay requirements},  number of relays are increased accordingly.  \amj{In this way, this scheme adds }    more  flexibility in \amj{a}  system to achieve  certain  QoS requirements.

\subsection{Optical Networks}
\label{subsec:optical-networks}
\amj{Concept of EC has also been used in OWC \cite{R148}.} OWC utilizes  ultraviolet, visible, and infrared light as  \amj{a} wireless medium to transmit  signal \cite{R11,R147}. Visibl light communications\footnote{for more details on achievable EC  in visible light communications, see Section \ref{sec:case-studies}.} is also a type of optical  \amj{communications}, that operates in  \amj{a} visible spectrum band (390-750nm). Extensive work has been done on optical wireless communications with    \amj{respect to} EC concept. Authors in \cite{R285},  have analysed  the performance  of MIMO antennas in optical communications with the help of the EC model. As compared to  traditional wireless networks, optical communications employ  turbulence fading channels to  \amj{accommodate}  fading conditions of optical environment. In this study, Gamma-Gamma turbulence fading condition with MIMO antennas to support  multiple users have been analysed  with the EC metric \amj{with stringent delay requirements}.  An  innovative and highly robust optical network framework named as petaweb has been proposed in \cite{R292,R294}. In this study, an in-depth analysis of optical networks with \blue{the} EC model has also been provided \amj{with the consideration of IP networks}.  \amj{In this work,} periodic fluctuations in   \amj{a} channel has been controlled through regulating  \amj{a} traffic flow.  
  
\subsection{WLAN}
\label{subsec:wlan}
Wireless local area network (WLAN) provides  \amj{a} connectivity of  two or more computers or devices over a limited area. WLAN follows the standard 802.11, and can  be connected to Internet through \amj{a} gateway \cite{R18,R60,R259}. EC concept in WLAN has  been used  to  \amj{investigate}  the performance of  WLAN  \cite{R146,R257,R258} \blue{for multimedia applications}. EC-based QoS analysis in WiFi networks has been discussed in  \cite{R121}. In this study, WiFi offloading with heterogeneous architecture has been explored with statistical QoS provisioning. An optimal resource allocation \amj{scheme}  (power allocation)  has been  \amj{developed using the EC model.}  For efficient estimation of  \amj{available} bandwidth in WLAN, WBest (a bandwidth estimation tool based on \blue{the  EC model} in WLAN) has been proposed in \cite{R260}. WBest operates with an  algorithm that comprises of two steps. In \blue{the}  first step,  achievable EC is estimated with the help  \blue{a} packet-pair approach, while the second step  \amj{provides the throughput analysis.}  This tool has been tested with many multimedia applications that demand higher bandwidth and stringent delay  requirement  for their transmission.

WLAN-based single-hop adhoc  networks with \blue{the  EC model}  has been discussed in \cite{R261}. In this work,  call admission control has been  \amj{investigated} with statistical delay guarantee. An efficient resource allocation algorithm has also been \amj{developed based on the EC model.}  Statistical QoS-provisioning in WLAN by considering  the EC model has also been studied in \cite{R306,R307}. In this study, 802.11 based mobile station is considered as a server. This server is then modeled as the Markovian bursty server. The known results or activity patterns from this server is then used to derive EC for  delay-sensitive and QoS-aware applications.  Proposed scheme has been tested through extensive simulations to validate the operation of the work. 

\subsection{Wireless Sensor Networks}
\label{subsec:sensor-networks}
Advances in micro-electro-mechanical systems (MEMS) technology has resulted into the wireless sensor networks (WSNs). WSNs now have found their applications ranging from civil, medical, to military   \cite{Rashid16jnca,R318,R319,R320}.  Limited battery life of tiny sensor nodes compel the researchers to come out with energy-conservation approaches while maximizing  network life time and throughput \cite{R175}.  Broad concept of EC has also been  used in WSNs for   \amj{investigating} delay and jitter. Performance of sensing operation and energy conservation approaches has been studied using   \blue{the  EC model} \amj{in} \cite{R228}.  Authors in \cite{R239}, have discussed  \amj{a}  wireless link  \amj{scheduling} approach with  achievable EC in WSNs.    Proposed scheduling approach assigns  time-slots to different users based on  \amj{source/arrival}  rate and \amj{required} delay constrained. Then  EC-based link-layer model is used to analyse this proposed scheduling approach to support  QoS-aware applications.  Proposed scheme has been compared with  traditional time-division multiple-access (TDMA) scheme and shows an improvement in throughput and energy-efficiency. Support of heterogeneous multimedia applications over cluster-based WSNs with \blue{the  EC model} has been studied in \cite{R265}. In this work, two-tier architecture for tiny sensing nodes has been proposed. In this two-tier architecture,  sensing nodes are provided with one antenna while the base stations (BS) are equipped with multiple antennas.   \amj{Low latency} applications are then sensed and transmitted over this architecture.  Performance of the network is then  \amj{investigated using the EC model}. 

\subsection{Mobile Wireless Networks}
\label{subsec:mobile-wireless-networks}

Mobility consideration in wireless networks demands  efficient management of wireless resources that should be  \amj{available} to all \amj{the } mobile users \cite{R262}. The main goal of any  \amj{mobile}  network is to enhance the flexibility and to reduce the cost of  \amj{a required} architectural \amj{layout}  \cite{R279,R280}. Efficient resource  \amj{management}  can be introduced in mobile networks and the performance of the network can also be  \amj{investigated}  using  \blue{the  EC model}. Authors in \cite{R282}, have provided a QoS-driven resource allocation scheme for  mobile wireless networks. In this work, the problem of power allocation and rate adaptation has been discussed with aim to maximize the throughput. This QoS-aware resource allocation  \amj{has been developed by considering the EC model.}

Cross-layer  \amj{resource allocation with \blue{the  EC model}}    to analyze the behavior of QoS-driven applications in mobile wireless networks has \blue{also} been discussed in \cite{R286}. In this cross-layer model,  MIMO antenna diversity has been modeled using  \amj{a} finite state Markove chain process, while the QoS-provisioning at  link-layer has been modeled using \blue{the} EC concept. This  scheme shows  \amj{an}  efficient interaction between the physical-layer and upper-layers while satisfying  \amj{a}  QoS-guarantee.  \amj{Mobility} issue with the EC modeling has also been  \amj{analyzed}  in mobile satellite networks \cite{R74}.  In this scheme, \blue{a}  statistical QoS-based power allocation has been  \blue{proposed}  while supporting  \amj{a}  high quality satellite wireless link.  \blue{In}  this approach,  physical wireless channel has been modeled as    \amj{a}  shadowed Rician model.   The  \amj{performance} of  \amj{a} proposed scheme  in supporting  QoS-aware  \amj{for} delay-sensitive applications has been   \amj{analyzed} using \blue{the  EC model}. \blue{An optimal resource allocation scheme to maximize the EC has been developed. The achievable EC has been studied with respect to different propagation conditions, delay exponent, and elavation angle. This study further confirms that, the achievable EC highly depends on the elevation angle in the mobile satellite networks.  }     
   
\subsection{Vehicular Adhoc Networks}
\label{subsec:vanet}
With the help of vehicular-adhoc Networks (VANETs),  dream of intelligent transportation in future smart cities can be realized \cite{R324}. However, this requires an intelligent management of different vehicular resources  \cite{R321,R249}. Different multimedia applications such as streaming videos in moving vehicles have \amj{also} been investigated with  EC concept  \cite{R238}. In this  research work, \blue{the}  packet-delivery \blue{ratio}  between the vehicle and road-side unit has been analyzed.   \blue{The  EC model} is  used to estimate the optimal distance between  \amj{a}  vehicle and road-side unit  \amj{and to investigate the transmission of}   QoS-aware  \amj{data}  with  \blue{loose}  delay  \blue{requirements}. \blue{Provisioning of low latency in vehicular communications with EC perspective has been studied in \cite{ner1}. The concept of latency violation probability (LVP) with optimal resource allocation such as power and spectrum has also been investigated. This EC-based latency analysis shows that the LVP can well be characterized with the notion of EC. By investigating the different latency requirements of safety-critical information, the sum rate of  vehicle-to-infrastructure (V2I) and vehicle-to-vehicle (V2V) links has been optimized by optimally allocating the resources, i.e., power and spectrum.}      

\subsection{Mesh Networks}
\label{subsec:mesh-Networks}
To minimize the dependency on one or more than one nodes for relaying the information to the destination, mesh networks have been proposed. In these networks,  participating nodes try to connect to as many as possible nodes for relaying their information to the destinations \cite{R266}. EC concept has also been utilized in this network topology with \amj{an} aim to analyse  the  \amj{required delay and QoS.}  In \cite{R276}, authors have proposed the wireless multi-hop mesh network to support  delay-sensitive applications. To   \amj{investigate} this mesh-topology,  the EC model   has been used. At the physical-layer, Rayleigh fading with fluid traffic model has been used. Extensive simulations have been performed under different traffic flows to validate the effectiveness of \amj{a}  proposed scheme. \blue{This study reveals that, in the scenario of multi-hop mesh networks, the route selection (routing path) is independent from the QoS requirements and other channel conditions. Furthermore, EC-based analysis of this multi-hop network shows that, the delay performance between the intermediate nodes is less sensitive to the arrival rate as compared to the end-to-end delay analysis.  }

\subsection{Cellular Networks}
\label{subsec:cellular}
\begin {figure*}
\footnotesize
\centering
\begin{tikzpicture}[
  level 1/.style={sibling distance=38mm},
  edge from parent/.style={->,draw},
  >=latex]

\node[root,minimum height=3em,text width=6cm,fill=blue!20] {Effective Capacity Model In Cellular Networks}
  child {node[level 2] (c1) {Long Term \\Evolution}}
  child {node[level 2] (c2) {Traditional Cellular\\ Networks    }}
  child {node[level 2] (c3) {Small  Cells \\ Networks  }}
  child {node[level 2] (c4) {Femto\\Cells }}
  child {node[level 2] (c5) {5G \\Networks }};

\begin{scope}[every node/.style={level 3}]
\node [below of = c1, xshift=10pt,yshift=-15pt] (c11) {Video Streaming \\ \cite{R33}  };
\node [below of = c11,yshift=-15pt] (c12) {Support of Real-Time Application Over LTE-Relay Networks \\ \cite{R167}};
\node [below of = c12,yshift=-15pt] (c13) {VoIP Application \\\cite{R176} };

\node [below of = c2,xshift=10pt,yshift=-15pt]	 (c21) {2D Applications \\\cite{R105}
};
\node [below of = c21,xshift=0pt,yshift=-15pt] (c22) {Multi-User QoS Provisioning \\\cite{R264}};
\node [below of = c22,,xshift=0pt,yshift=-15pt] (c23) { Downlink Scheduling of Real-Time Traffic\\ \cite{R291}};
\node [below of = c23,,xshift=-0pt,yshift=-22pt] (c24) { QoS Provisioning In Cellular D2D Communications\\ \cite{R151}}; 
\node [below of = c3, xshift=00pt,yshift=-28pt] (c31) {Support for Real-Time Applications in Heterogeneous Cells \\\cite{R68} };
\node [below of = c31, yshift=-30pt] (c32) {QoS-Provisioning in Visible Light supportive Cellular Networks \\ \cite{R117} };
\node [below of = c4, xshift=10pt,yshift=-15pt] (c41) {Video Streaming \\ \cite{R140} };
\node [below of = c41,yshift=-15pt] (c42) {Energy-Efficient Based QoS Provisioning\\ \cite{R67,R199}};
\node [below of = c42,yshift=-18pt] (c43) {QoS-Provisioning In OFDMA-Based Femto-Cells \\\cite{R186} };
\node [below of = c43,yshift=-22pt] (c44) {QoS-Provisioning In Two-Tier  Femto-Cells\\\cite{R160,R177,R190,R191} };
\node [below of = c5, xshift=10pt,yshift=-15pt] (c51) {Video Streaming \\\cite{R122,R125} };
\node [below of = c51,yshift=-15pt] (c52) {Uplink QoS Provisioning \\\cite{R129}};
\node [below of = c52,yshift=-17pt] (c53) {Energy-Efficient QoS Provisioning In Green 5G Networks \\ \cite{R138} };
\node [below of = c53,yshift=-22pt] (c54) {QoS-Provisioning in Full-Duplex 5G Networks \\ \cite{R182} };

\end{scope}

\foreach \value in {1,2,3}
  \draw[->] (c1.180) |- (c1\value.west);

\foreach \value in {1,...,4}
  \draw[->] (c2.180) |- (c2\value.west);

\foreach \value in {1,...,2}
   \draw[->] (c3.180) |- (c3\value.west);
   
\foreach \value in {1,2,3,4}
  \draw[->] (c4.180) |- (c4\value.west);
  
\foreach \value in {1,2,3,4}
  \draw[->] (c5.180) |- (c5\value.west);
   
\end{tikzpicture}
\caption {Effective capacity measurements in cellular networks can be classified into traditional cellular networks, small Cells, Femto-cells, LTE, and 5G Networks.  }
\label{fig:celluar-ec}
\end{figure*}
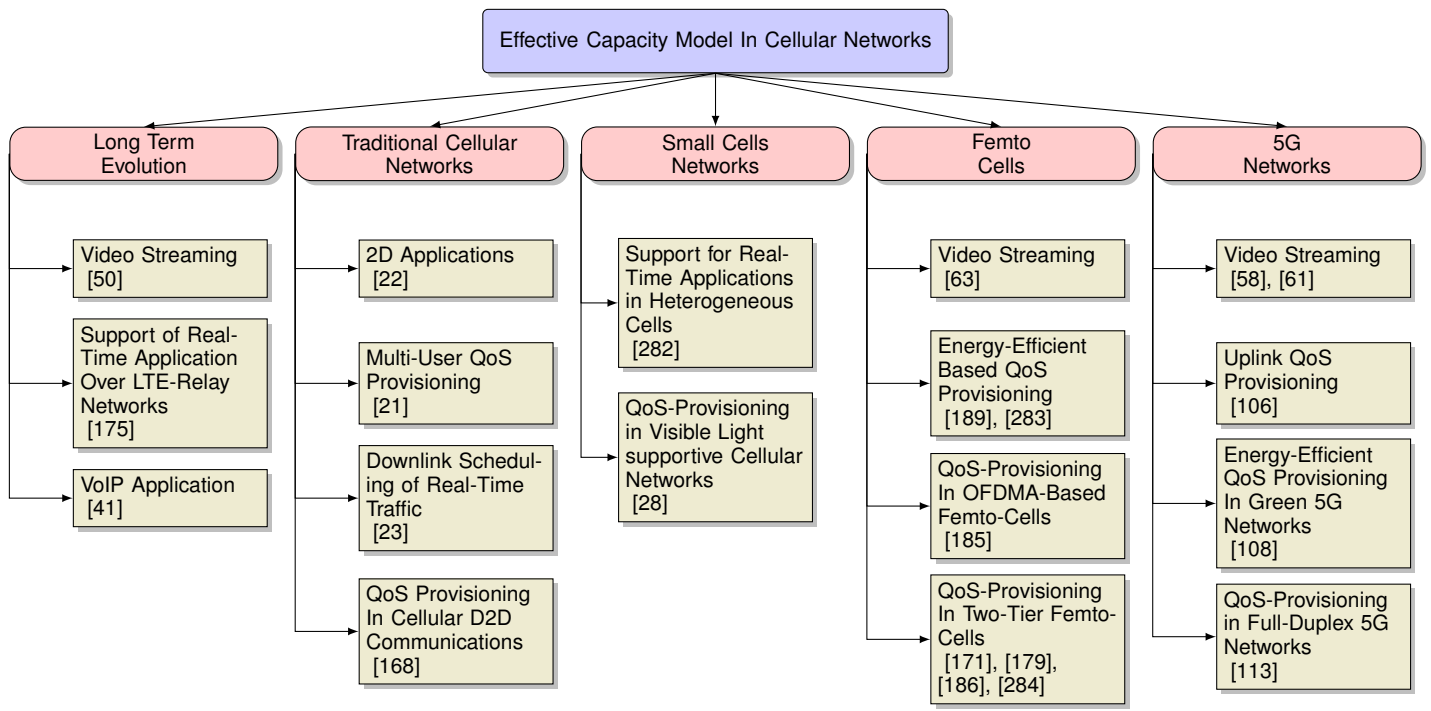

\amj{Performance of cellular networks can also be analysed with the help of \blue{the  EC model} while considering delay constraints. Study of cellular networks with \blue{the  EC model} has been classified into LTE, small cells, femto cells, and 5G networks, as has been shown in Figure \ref{fig:celluar-ec}.}

\subsubsection{Traditional Cellular Networks}
Advances in telecommunication systems have resulted into the emergence of  advance cellular networks such as LTE and 5G. In cellular networks,  area for which the radio connectivity is to be provided is usually divided into  cells with hexagonal or with some other shape according to terrain and land characterization \cite{R105}. With  \amj{an}  increase in  cellular users,  demand for advance applications such as multimedia has also been increased. To guarantee   \amj{statistical} QoS provisioning in cellular networks,  EC-based modeling of cellular radio    has    been performed  \blue{in} \cite{R151}.   \blue{In this work, the achievable EC of  D2D communications underlaying cellular networks has been investigated in detail. A constraint optimization problem  that can maximize the cellular users EC has been formulated. This study reveals that as the arrival rate of  D2D pairs increases, the EC decreases and interference to the BS increases. This framework is then used to develop an adaptive power allocation scheme.}  Authors in \cite{R264}, have used  \blue{the  EC model} to  \amj{investigate}  the  optimal resource allocation scheme  for downlink in cellular networks. In this study,  multiuser \amj{statistical} QoS-provisioning in downlink cellular  networks has been \amj{taken into consideration.}    Problem of downlink resource allocation has been formulated with power and delay constraints into optimization problem and then solved using two-steps procedure. \blue{In this work,  EC of  downlink users with adaptive power allocation scheme has been compared with  equal-length time division (TD) approach. The proposed scheme outperforms the equal-length TD with respect to the achievable EC of two users.   } 
\subsubsection{Femto Cells}
To extend the coverage of existing cellular networks,  concept of femto cells  has been proposed \blue{in}  \cite{R67,R140,R198}. At the edge of cellular networks, where the coverage is unavailable or limited,  femto-cell technology is usually used. It can  \amj{accommodate} approximately eight to ten users depending upon the infrastructure used.  The    \blue{performance}  of femto-cells have  been studied with the EC model \blue{in}  \cite{R160,R177,R199}.  In \cite{R186},  \amj{downlink} resource allocation for femto cells have been studied with the EC model.  Statistical QoS provisioning has been ensured with \blue{a} minimum energy consumption. In this OFDMA-based resource allocation scheme,  channel and power allocation have also been analysed  \blue{using the EC model} for  downlink femto cells. Authors in \cite{R190}, have come out with \blue{an}  energy-efficient    QoS-provisioning   \blue{method based on the} EC concept in uplink femto-cells. \blue{A}  two tier-architecture consisting of macro-users and femto-cells \amj{has been proposed in this study.}  Interaction between  two users for  accessing   \amj{random} resources and supporting QoS-aware  applications has been addressed with  \amj{a}  Q-learning approach. The proposed scheme has been \amj{investigated} with \blue{the  EC model} and shows  improved performance in terms of convergence speed. Energy-efficiency with statistical QoS-provisioning in two-tier femtocells have also been analysed with \blue{the  EC model} in \cite{R191}. In this scheme, \blue{a}  price-based power control policy has been explored to  mitigate  inter-cell interference.   Problem of  power allocation  has been formulated as  \amj{a}  Nash equilibrium and  is solved using  \amj{a}  particle swarm  optimization scheme.  \blue{The} EC-based delay and performance analysis   \blue{helps to validate}  the effectiveness of  \blue{the}  proposed scheme.   

\subsubsection{5G and Byond 5G Networks}
5th generation (5G) is the next-generation  \amj{wireless networks}  that    promises higher bandwidth, lower-latency, \amj{ultra-reliability},  and other features such as lower energy-consumption \cite{R122}. With higher data rates, 5G networks will have the flexibility to meet the needs of future implementation of Internet of Things (IoT)~\cite{Khan17wcom}.  Higher data rates in 5G networks  \amj{can be} provided with the help of advance technologies such as FD communications, mmWave, and NOMA  \cite{R125}.  \amj{Achievable EC of 5G networks  \blue{is used}  to investigate the  \blue{performance of low-latency applications} such as URLLC \cite{urllc1,R129}.}  Green 5G wireless mobile networks with  \blue{the  EC model} has \blue{also} been discussed in \cite{R138}. In this study, \blue{a} statistical QoS-driven power allocation in 5G networks  \blue{with SISO and MIMO antennas} has been discussed.  Concept of effective power efficiency (EPE) has been used  in conjunction with the \amj{achievable EC.}  EPE for the SISO and MIMO antennas has been  \amj{investigated} through extensive simulations  involving different delay bounds. 5G mobile networks with EC concept  to analyse the  performance of  heterogeneous network resources  has also been discussed in \cite{R182}. In this scheme, for the first time  channel coupling has been  \amj{studied} using  \blue{the  EC model}. In this study, to thoroughly understand the impact of channel coupling on  network performance,   three different case studies, i.e., FD communications, CRNs, and D2D communications, based on \blue{the  EC model}     have been \amj{presented.} \blue{In this EC-based diverse delay analysis of different technologies, the heterogeneous delay bounded proposed achitecture has been compared with the traditional homogeneous delay bound QoS schemes. The achievable EC of the proposed architecture outperforms the traditional homogeneous delay bound QoS schemes. Furthermore, different delay bounds of the BS depending on various enabling technologies of 5G has been chalked out.}

\subsubsection{Small Cells}
To increase  \amj{a} cellular network capacity, flexibility, and resiliency,  concept of small cells have been envisioned as the extension to  long-term evolution (LTE). Small cells can reuse the existing licensed and unlicensed spectral resources, hence can  \blue{improve}  the usage of spectral resources \cite{R117}. Small cells can include  femto, macro, and micro-cells. EC concept has  been used under the small cells umbrella.  In \cite{R68}, the installation of few small cells (from 1 to 3) \amj{in} heterogeneous networks environment has been analyzed using \blue{the  EC model}.  In this study,  EC relief option related to analysing,  planning, and management of small cells have been \amj{investigated in detail}. \blue{EC analysis of this relief option shows that,  properly deployed fewer small cells (1 to 3) can enhanced the network capacity three times. This deployment will also enhance the non-uniformity in traffic load.}
\subsubsection{LTE}
3rd generation partnership project (3GPP) has introduced the long-term evolution (LTE) that is based on  existing GSM and UMTS standards \cite{R290,R293}. LTE with its new release  called as 4th generation (4G) can promise higher data rates and capacity for mobile users \cite{R33}. Performance of LTE networks has also been tested using the concept of EC metric  \cite{R167}.  Air-interface in LTE networks can support  delay-sensitive and bandwidth hungry applications. Cross-layer scheduling scheme involving  EC-based modeling of link-layer in LTE has been studied in \cite{R176}. In this downlink LTE networks,  QoS-driven energy optimization has been analysed with \blue{the  EC model}. Extensive simulations of the proposed scheduling scheme shows  48\% of reduction in total energy-consumption as compared to other traditional LTE frameworks.   

\subsection{Summary and Insights}
In this section, different wireless networks such as CRNs, cooperative networks, optical networks, and cellular networks with \amj{their achievable EC are  \blue{surveyed}  in detail. Maximization in EC in different networks are studied to understand the QoS provisioning for delay-sensitive applications. This will also be a corner stone to study the future networks with stringent latency requirements.  Depending upon the white space utilization, CRNs can be categorized into underlay, overlay, and interweave CRNs. However, it is a challenging task to determine which technique is suitable for applications with stringent  \blue{or}   \blue{loose}  delay constraints. \blue{The  EC model} can be a helpful tool to explore this issue further. In  CRNs, another challenge is to find out the quantification between different techniques such as delay, reliability, spectrum sensing, PUs activity, and spectrum mobility. In this regard,  using \blue{the} EC metric can simplify the performance analysis of CRNs to understand the relation/trade-off between different techniques of spectrum sensing, PUs activity, delay, and others.}

\amj{In cooperative networks, 70\% of the delay and energy-depletion is due to the relays and BS's. There exists a serious challenge to understand the performance of   cooperative networks with delay-constraints.  To address this challenge, EC provides a clean slate and is considered as an accurate and flexible tool to understand performance of cooperative networks. 
}
\begin{table*}
\scriptsize
\centering
\caption{Achievable  Effective Capacity  in FD communications can be studied while considering  different SIS approaches, network types, fading channels, and antenna requirements.}
\label{tab:wmcrns-applications}
\begin{tabular}{|p{2cm}|p{2cm}|p{1cm}|p{3cm}|p{3cm}|p{2cm}|}
\hline

 \multicolumn{1}{|c}{\bf SIS Approaches} &   &   \bf Study&\bf Network Type&\bf Fading Channel Used & \bf Antenna Design    \\\hline
 Active SIS Approaches& Digital SIS Approach&\cite{R156}&FD-Relay Networks& Rayleigh Fading Channel  &Not Defined\\
\cline{2-6}
& Analog SIS Approach&\cite{R136}&FD-Relay Networks &Rayleigh Fading Channel & Two Antennas\\\cline{2-6}
&Analog and Digital SIS Approach&\cite{R21}&FD-Relay Networks&Rayleigh Fading Channel& Single Antenna\\\cline{3-6}
&&\cite{R133} &FD-Relay Networks&Nakagami-$m$ Fading Channel &Single Antenna\\\cline{3-6}
&&\cite{R134}&FD-Cognitive Radio Networks&Nakagami-$m$ Fading Channel&Not Defined\\\cline{3-6}
 &&\cite{R143}&FD-Relay Networks&Rayleigh Fading Channel & Not Defined \\
\cline{3-6}
& &\cite{R312}&FD-Relay Networks &Rayleigh Fading Channel &Not Defined  \\\hline
\multicolumn{2}{|c|}{Passive SIS Approaches}&\cite{R64}&FD-Relay Networks& Rayleigh Fading Channel &Directional Antenna \\\cline{3-6}
\multicolumn{2}{|c|}{}&\cite{R123}&FD-Relay Networks&Rayleigh Fading Channel &Single Antenna\\\cline{3-6}
\multicolumn{2}{|c|}{}&\cite{R125}&FD-Cellular Networks&Nakagami-$m$ Fading Channel &MIMO Antenna\\\cline{3-6}
\multicolumn{2}{|c|}{}&\cite{R182}&FD-Cellular Networks&Nakagami-$m$ Fading Channel &MIMO Antenna\\\cline{3-6}
\multicolumn{2}{|c|}{}&\cite{R218}&FD-Relay Networks&Nakagami-$m$ Fading Channel&Bidirectional Antenna\\\cline{3-6}
\multicolumn{2}{|c|}{}&\cite{R308}&FD-Relay Networks&Rayleigh Fading Channel&Not Defined\\\cline{3-6}
\multicolumn{2}{|c|}{}&\cite{R309}&FD-Relay Networks&Rayleigh Fading Channel &MIMO Antenna\\\cline{3-6}
\multicolumn{2}{|c|}{}&\cite{R311}&FD-Relay Networks&Rayleigh Fading Channel &MIMO Antenna \\\hline
\multicolumn{2}{|c|}{Hybrid SIS Approaches}&\cite{R221}&FD-Relay Networks&Rayleigh Fading Channel &Two Antennas\\\cline{3-6}
\multicolumn{2}{|c|}{}&\cite{R310}&FD-Cellular Networks&Rayleigh Fading Channel&Omni-directional Antenna \\\hline
\end{tabular}
\label{table:sis} 
\end{table*}

\section{Effective Capacity for Full-Duplex Communications}
\label{sec:full-duplex}
Full-duplex (FD) communication has gained much attention due to the provision of higher data rates \cite{R125}.  QoS-aware multimedia applications are often regarded as  bandwidth hungry. Therefore,  concept of FD communications has been introduced to support the ballooning demand of real-time and  delay-sensitive applications. However,  dream of simultaneous transmission and reception (FD communications)   can only be realized while suppressing  self-interference (SI). 

Concept of link-layer channel model, i.e., EC   for analysing the performance of FD communications has been extensively explored in  \amj{existing}   literature. EC-based delay analysis with  half-duplex communications design has been provided in detail in \cite{R44,R48,R72}. In comparison to  half-duplex communications, FD design can theoretically double  data rate.  Potential advantages of FD communication can only be achieved with the implementation of proper self-interference suppression (SIS) approaches. 

  \amj{In this Section,}  we have considered \amj{the achievable EC of}   FD communication with different SIS approaches   \blue{under}  delay-outage probability \blue{constraints}. Self-interference in FD communications can be minimized with the help of either passive or active SIS approaches. In passive SIS approach,  SI is mitigated with antenna and signal propagation approaches, such as antenna shielding, antenna separation, beamforming, and antenna polarization effects. In active SIS approach,  radio frequency (RF) canceller and baseband canceller are utilized to achieve  simultaneous transmission and reception.  Table \ref{table:sis} shows  \amj{a}  comparative view of different SIS approaches while highlighting the fading channel employed and antenna design involved for achieving  \amj{a}  reliable SIS approach. Below is the description of  \amj{work with achievable EC of FD communications}  while considering  passive and active SIS approaches. 

\subsection{Passive Self-Interference Suppression Approach}
\label{subsec:passive}
To mitigate  SI through passive SIS approach,  signals are treated with some antenna separation, shielding, and polarization before the signal actually enters   \amj{a} local transmitter.  In \cite{R64}, authors  \amj{have analysed the   arrival/source} rate of FD two hop networks using \blue{the}  EC concept \amj{with certain delay-bound}.  SI is minimized with the help of passive SIS approach  \amj{while taking into consideration} various buffer constraints (with respect to arrival rates)  at relay, source, and destination node.   Performance of the proposed scheme has been tested  with  \amj{achievable } EC  and shows improvement as compared to other \amj{state-of-the-art} schemes.   Time-critical and delay-sensitive applications such as video transmission with FD communications is investigated in \cite{R123},  while considering  \blue{the  EC model}.  To predict the quality of the video streams, EC metric is used for detailed analysis with passive SIS approach.   Passive SIS approach with  delay constraints in this FD communications is compared \amj{with}  other two sub-optimal strategies.

Statistical  QoS provisioning with \blue{the  EC model}   \amj{has also been}  studied in emerging wireless networks such as \amj{FD-enabled} 5G networks \cite{R125}. In this study,  passive SIS approach with MIMO antennas is used to suppress the SI by adjusting the transmission power.  With  passive SIS approach, Quadrature-OFDMA (Q-OFDMA) scheme with D2D communications has  been analyzed with EC \amj{under stringent delay requirements}.   \amj{In this study, the maximization in EC has been studied under Nakagami-$m$ fading conditions.}      Authors in \cite{R182}, have also used the concept of EC with \amj{FD-enabled 5G}  networks while using  passive SIS approach. In this scheme,  heterogeneous statistical QoS provisioning in 5G networks has been studied based on three network architectures namely, FD architecture, CRNs, and D2D network.

EC-based delay analysis with advance passive SIS approaches such as  local transmit power unrelated self-interference (LTPUS) and local transmit power related self-interference (LTPRS) in FD communications have been discussed in \cite{R218}. In LTPUS,  SIS approach does not directly depend on  \amj{a}  power level, however in LTPRS the power of  local transmitter is also considered to mitigate  \amj{an}  excessive SI.  Arrival rate in conjunction  with  SIS approach \amj{with required delay-outage probability}  has been analyzed using  EC  \amj{concept}.  Cross-layer  \amj{resource allocation}   while using the SIS approach  in FD communications \amj{with achievable EC}  has been discussed  in \cite{R308}. In this study,  SI has been overcome by optimally controlling  power and \amj{by} using  the efficient  relaying scheme. AF  method of relay selection with EC concept has been analysed in detail.  Proposed scheme has also been  evaluated through extensive simulations and compared with  \amj{a} traditional FD scheme with direct transmission. EC-maximization  shows that,  proposed work shows two-fold improvement in throughput as compared to the other \amj{state-of-the-art} work. 

Statistical QoS provisioning with passive SIS approach based on buffer-aided  relaying method in FD relay networks has been explored in \cite{R309}.  SI has been controlled by optimally controlling  transmit power of  \amj{a}  local transmitter.  \amj{EC framework in this scheme has been used to investigate the resource allocation scheme for optimally assigning the  power. }    Relay-mode selection criteria while utilizing the two-way MIMO systems with EC   has been researched in \cite{R311}. Throughput under various QoS constraints and optimal selection of half-duplex or FD mode has been analysed with  \amj{achievable} EC.   Proposed scheme shows better performance at low signal-to-noise ratio (SNR) with FD mode. However, at high SNR,   half-duplex scheme outperforms the FD mode. These comparisons have been made while employing  \blue{the  EC model}  for FD mode. 
\subsection{Active Self-Interference Suppression Approach}
\label{subsec:active}
Baseband canceller and RF canceller are used to actively mitigate  SI at  \amj{a}  local transmitter and receiver.  Active SIS approaches   in FD communications \amj{with EC}  has been  discussed in literature. Through  \amj{a}  proper implementation   of active SIS approach, 40-50 dB of  SI can be reduced.   \amj{Achievable EC}   in FD communications while considering  active SIS approach can be studied by classifying the active SIS approaches into digital, analog, and combination of these approaches. 
\subsubsection{Digital Self-Interference Suppression Approach}
Non-linearities in  analog-to-digital converter (ADC) and irregularities in  \amj{an} oscillator  \amj{can} result into  SI.  SI resulting from such  \amj{factors}  can be mitigated with the help of digital SIS approach. Dynamic range of  \amj{a} receiver ADC, can also be handled with the digital SIS approach. Authors in \cite{R156}, have used the concept of EC to understand and  \amj{model}  the digital SIS approach in two hop networks with FD relays. In this work,    decode-and-forward (DF) relaying method has been used to support the cooperative communications. Both the source and relay queues, \amj{with certain delay-bound}  have been  \amj{investigated}  with \blue{the  EC model}. A trade-off between source and relay queues regarding  statistical QoS provisioning has been achieved while actively suppressing  SI at relay node.   \amj{Simulations have been performed to validate the EC-based mathematical framework for this SIS approach for FD networks.}  \blue{The EC analysis shows that the tradeoff between the two queues improves the sum of the effective capacity. This framwork also provides another insights regarding the importance of buffers at the relay. The relays with buffer shows better performance as compared to the relays without buffers.}

\subsubsection{Analog Self-Interference Suppression Approach}
 Complexities at  ADC can also introduce SI. This SI can be effectively suppressed with the help of analog SIS approach. Various techniques such as time-domain algorithms, sequence-based methods or adaptive interference suppression have been introduced to actively suppress the SI.      Analog SIS approach with optimal power allocation under channel uncertainties while using the concept of EC  has been discussed in \cite{R136}. In this work, not only  SI is suppressed but also the loop-interference has  been  minimized  \amj{with stringent delay requirement.}    Problem of SI, loop interference, and optimal power allocation has been formulated by employing  Taylor optimization and solved using  Lagrange dual approach.  Proposed scheme has been  \amj{investigated}   with EC    and shows improved performance \amj{as compared to the other state-of-the-art schemes}  while residing within statistical delay QoS constraints.   

\subsubsection{Analog and Digital Self-Interference Suppression Approach}
Combination of analog and digital SIS approach is also used to mitigate  SI that \amj{can} result from the complexities, non-linearities in ADC, and irregularities in oscillator. This analog and digital SIS approach  \amj{with achievable EC} has also been \amj{investigated in the existing work}.    Analog and digital SIS approach in FD relay networks with buffer-aided relaying scheme \amj{with EC} has been   \amj{discussed in } \cite{R21}.  Infinite size queues at   \amj{a} source and a relay node has been taken into consideration with Rayleigh fading channel. \amj{In this work, EC is used to find the arrival/source rate, and depending upon the source rate, the relaying is performed, thats why this scheme has been named as ``selection relaying''. }  

 Achievable EC with analog and digital SIS approach has also been studied  \amj{with} heterogeneous QoS requirements \cite{R133}. In this FD-relay networks,  DF relaying scheme with Nakagami-$m$ fading channel has been  \amj{investigated}  with EC.  \amj{In this work,}  heterogeneous QoS-aware resource allocation \amj{scheme} has  been  \amj{developed}  with the help of EC.   \amj{Concept of} EC  with \amj{both} analog and digital SIS approach has also been explored in FD-CRNs \cite{R134}. In this scheme,  simultaneous spectrum sensing and transmission in CRNs in conjunction with  statistical QoS provisioning has been   \amj{modeled}  with EC. In this scheme,  Nakagami-$m$ fading channel has been used   and  probabilities of false alarm and miss detection are derived. An in-depth   EC-based QoS analysis with  proposed FD-CRNs has been carried out to study the  \amj{required delay-violation probability}  for  real-time applications.

Another buffer-aided relay selection scheme in FD-relay networks with analog and digital SIS approach has been  \amj{explored in detail}  using  EC concept in \cite{R143}. A trade-off between  statistical delay constraints on two concatenated queues has been derived.  \amj{Then a maximum constant arrival rate is derived using the EC model.}      SI at the FD-relay has been minimized while considering  analog and digital SIS approach with Rayleigh fading channel. EC-based performance analysis shows that proposed scheme shows an improved throughput as compared to  other \amj{state-of-the-art} schemes.   \amj{Achievable}  EC  with analog and digital SIS approach has also been studied in FD-relay networks with AF relaying protocol \cite{R312}. This work provides the idea of better link quality \amj{selection} between   \amj{a} source and relay \amj{node} to support the desired QoS requirements for  FD-relay networks.  Link quality between   \amj{a} source and relay node has been  \amj{investigated}  while considering the  EC \amj{model}. This  analysis shows that the SIS approaches \amj{also}    affect the link quality in FD relay networks.                                 

\subsection{Hybrid Self-Interference Suppression Approach}
\label{subsec:hybrid}
 \amj{In some cases,}  both  passive and active SIS approaches are utilized to completely suppress  SI.  With  hybrid SIS approach,  concept of EC has also been   \amj{employed to investigate the proposed SI approach.}   Authors in \cite{R221}, have used the concept of EC for FD-relay networks with hybrid SIS approach. In this study,   both  AF and DF relaying schemes with their FD and half-duplex support have been  \amj{discussed in detail.}   Further,  a new control factor named as the cancellation coefficient has been proposed and both the half-duplex and FD mode has been analyzed with   \amj{achievable} EC.  Extensive analysis based on EC concept reveals that,  hybrid mode comprising of half-duplex and FD operation shows better performance. Hybrid SIS approach with \blue{the  EC model} has also been studied in FD-cellular networks \cite{R310}.  SI resulting from  omni-directional antenna has been mitigated with the help of passive and analog SIS approach.  Whole network with  hybrid SIS approach  has been modeled as  \amj{a} Matern point process  and then is analysed with \blue{the} EC concept.

\subsection{Summary and Insights}
\amj{Achievable EC in FD-communications with SIS approaches has been reviewed in this section. Simultaneous transmission and reception (FD communications) with stringent QoS provisioning is a challenging task.  \blue{The EC based analsyis of FD-relays provides important insight regarding the provision of buffer at the relay. The FD relays with buffers shows improved performance as compared to the relays without buffers.  }  Investigating the achievable EC of FD communications with other networks such as CRNs and  cellular networks further adds the complexity. Taking the closed-form expression of EC with FD communications is more complex than ergodic and Shannon capacity. Designing of  resource allocation schemes with proper SIS approaches for FD communications invites the future researchers to explore this dimension with achievable EC.} 
 \begin {figure*}
\footnotesize
\centering
\begin{tikzpicture}[
  level 1/.style={sibling distance=100mm},
  edge from parent/.style={->,draw},
  >=latex]

\node[root,minimum height=3em,text width=6cm,fill=blue!20] { EC-Based Performance Analysis for Retransmission Schemes    \\Sec \ref{sec:retransmission}}
  child {node[level 2] (c1) { ARQ\\Sec \ref{subsec:arq} }}
  child {node[level 2] (c2) {Hybrid-ARQ (HARQ)\\Sec \ref{subsec:harq}  }};
\begin{scope}[every node/.style={level 3}]
\node [below of = c1, xshift=105pt,yshift=-2pt] (c11) { Traditional ARQ \\\cite{R219} };
\node [below of = c11,xshift=-90pt,yshift=10pt] (c12) {Selective Repeat ARQ (SR-ARQ) \\\cite{R91} };
\node [below of = c12,xshift=90pt,yshift=6pt] (c13) {Network Coded ARQ (NC-ARQ)  \\\cite{R06}};
\node [below of = c13,xshift=-90pt,yshift=6pt] (c14) { Two-Mode ARQ\\\cite{R06} };

\node [below of = c14,xshift=90pt,yshift=6pt] (c15) { Multi-Layer ARQ\\\cite{R06} };
\node [below of = c2, xshift=105pt,yshift=-2pt] (c21) { Traditional HARQ\\\cite{R10} };
\node [below of = c21,xshift=-90pt,yshift=10pt] (c22) {Type-1 HARQ\\\cite{R165} };
\node [below of = c22,xshift=90pt,yshift=6pt] (c23) {Chase combining HARQ (CC-HARQ) \\\cite{R165}};
\node [below of = c23,xshift=-90pt,yshift=6pt] (c24) {Incremental redundancy HARQ (HARQ-IR) \\\cite{R165,R247,R313} };
\node [below of = c24,xshift=90pt,yshift=6pt] (c25) {Persistent HARQ\\\cite{R06} };
\node [below of = c25,xshift=-90pt,yshift=6pt] (c26) {Truncated HARQ\\\cite{R06} };
\end{scope}

\foreach \value in {1,...,5}
  \draw[->] (c1.180) |- (c1\value.west);

\foreach \value in {1,...,6}
  \draw[->] (c2.180) |- (c2\value.west);

\end{tikzpicture}
\caption { Existing work on effective capacity while employing  retransmission schemes are based on different flavors of ARQ and Hybrid-ARQ (HARQ).}
\label{fig:retransmission}
\end{figure*}
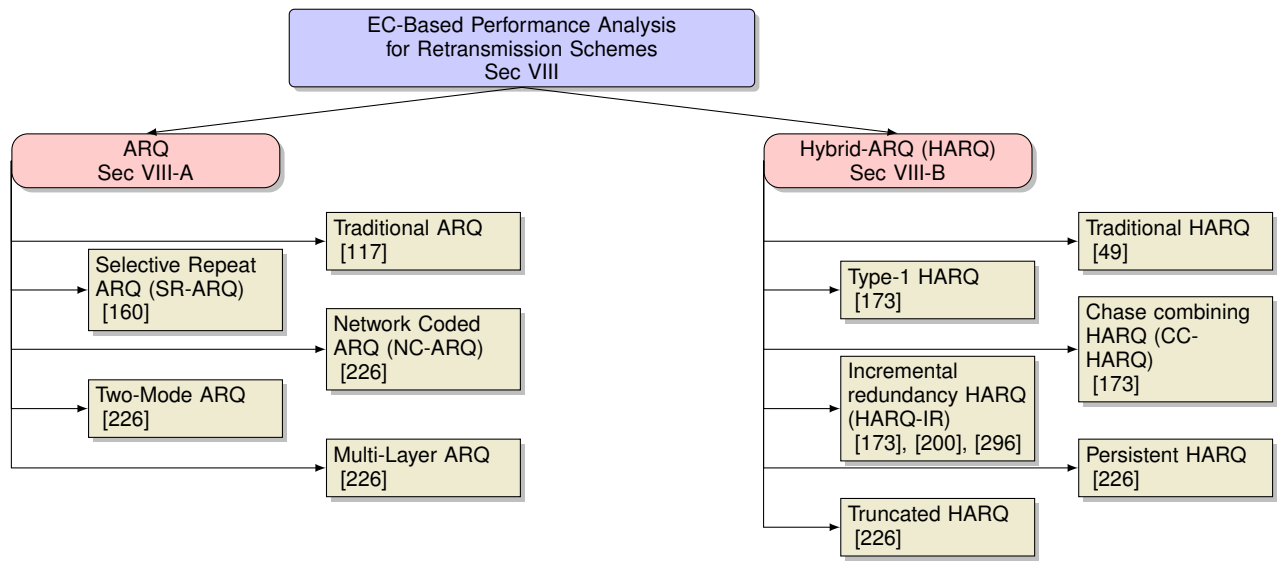

%
%
\section{Effective Capacity and Re-Transmission Schemes}
\label{sec:retransmission}

Packet switched networks are often prone to packet loss. Therefore,  packet retransmission schemes  \amj{are}   employed to achieve  reliable communications in  these networks.  Two versions of packet retransmission schemes, namely, automatic repeat request (ARQ) and hybrid automatic repeat request (HARQ), are extensively used to achieve  \amj{a}  better reliability in the networks. These retransmission schemes  \amj{use}  different forms of acknowledgement \amj{mechanisms}  to recover the loss packets in \amj{a} network.    \amj{Achievable EC for the various retransmission schemes has  been discussed in the existing literature.}    In  \amj{this}  paper,   current work on retransmission schemes   \amj{with achievable} EC  has been classified into ARQ and HARQ. Figure \ref{fig:retransmission}, shows  \amj{a} classification of  existing work on retransmission schemes that use  EC \amj{concept}. Different versions of ARQ and HARQ \amj{with  EC  } have also been  \amj{surveyed in this Section.}    Below is the description of  retransmission schemes that considers  the EC model.      

\subsection{Automatic Repeat Request (ARQ)}
\label{subsec:arq}
ARQ is extensively used to achieve the \amj{required}  reliability in  networks.  Concept of EC in conjunction with  ARQ has been  used to anaylse the  statistical delay provisioning    while residing within certain QoS constraints. The achievable EC of   various versions of ARQ ranging from traditional to two-mode ARQ have also  been \amj{explored}.  Details of some of these versions  \amj{are}  provided below:

\subsubsection{Traditional ARQ}
Authors in \cite{R219}, have    \amj{discussed the EC maximization  with}   adaptive modulation and coding (AMC) and ARQ schemes in underlay CRNs. \amj{In this study,} an optimal power allocation scheme has been discussed while meeting the stringent requirements of  packet-error rate (PER) and delay constraints. Power requirements and throughput were also analysed by using  \blue{the  EC model}.      

\subsubsection{Selective Repeat ARQ (SR-ARQ)}
ARQ with  finite \amj{retranmission} persistence \amj{scheme} such as truncated ARQ are utilized to \amj{recover the lost packets in lossy links.}     However, as compared to  finite \amj{retranmission} persistence,  ARQ with infinite persistence  \amj{can}  also be used to    \amj{achieve the required}  reliability in  network. The achievable EC of ARQ methods with  infinite persistent named as selective repeat ARQ (SR-ARQ)  has been analysed  in    \cite{R91}. In this work,   \amj{performance} of SR-ARQ has been analyzed in two approaches. First,  \amj{a} discrete-time  Markov chain (DTMC) model is used to analyze the queuing behavior and arrival process. In the second approach, concept of effective bandwidth and EC are used  \amj{to find the maximum arrival/source rate with desired delay-bound}  with infinite persistent SR-ARQ approach. 
\subsubsection{Network Coded ARQ (NC-ARQ)}
 Concept of network coding (NC) has been extensively used to \amj{optimize the flow of digital data in a network.}   To communicate  multiple packets over the same time \amj{slot},  concept of network-coded ARQ  (NC-ARQ) has been proposed. Authors in \cite{R06}, have provided a comparative view of different retransmissions schemes while considering the concept of EC  model. In this study,  NC-ARQ has been proposed and compared with  other versions of ARQ and HARQ. To achieve the reliability in  transmission, multiple packets are transmitted in a single stream.  Performance analysis of NC-ARQ based on  \amj{achievable EC},  shows that NC-ARQ outperforms the other \amj{state-of-the-art} retransmission schemes. 
 
\subsubsection{Two-Mode ARQ}
In \cite{R06},  \amj{the maximization of EC with}  two-mode ARQ in comparison to  traditional ARQ with  Gilbert-Elliot block fading channel has been \amj{investigated in detail}.    In this two-states or two-mode ARQ,  \amj{a} channel always assumes either of the two states \amj{and these two states at any time can be,}  bad-to-good, bad-to-bad, good-to-bad, \amj{and}  good-to-good. During, each mode or state, \amj{achievable EC has been investigated for the proposed ARQ with various retransmission parameters.}   
\subsubsection{Multi-Layer ARQ}
Similar to the idea of NC-ARQ, multi-layer ARQ can also transmit multiple packets in a single transmission \cite{R06}. By adjusting the transmission power,  \amj{various} codewords are communicated in a single stream. Receivers are intelligently designed to decode these codewords by considering the channel fading conditions.   Performance of  multi-layer ARQ has been anaylsed with EC  \amj{to find the maximum arrival/source rate with given delay requirement.}   
\subsection{Hybrid Automatic Repeat Request \amj{ (HARQ)}}
\label{subsec:harq}
In wireless communications,   reliability  has been further strengthen with the introduction of  \amj{HARQ}. In HARQ,  traditional ARQ error-control \amj{mechanism},  has been \amj{further}  enhanced with the inclusion of high rate forward error correction (FEC) code. With the help of link-layer capacity model, i.e., EC, new versions of HARQ has also been analyzed  while residing within certain delay constraints. Below is the description of  various versions of HARQ in conjunctions with   EC model.     
\subsubsection{Traditional HARQ}
To deal with  challenging conditions of  \amj{dynamic}  wireless medium, different transmission schemes including the adaptive modulation coding (AMC) and HARQ have been introduced and then  \amj{have been investigated with their achievable EC.}  Authors in \cite{R10}, have provided a comparative view of AMC and HARQ while considering the EC model.  Delay caused by AMC and HARQ has been taken into account to analyse the performance of the above  both.  Performance analysis based on EC \amj{maximization} for HARQ and AMC shows that their performance decreases at  high SNR.  
\subsubsection{Type-I HARQ}
Type-I HARQ, is the simplest version of HARQ \cite{R165}. In type-I HARQ, during  transmission \amj{of packets},  error-detection (ED) and FEC \amj{mechanisms} are added to each message to make  transmissions \amj{of packets} more reliable. In \cite{R165}, type-I HARQ has been analysed with EC, while considering  different transmission rates and \amj{delay-constraints}.  In this retransmission scheme, data packets are encoded into one codeword by using one of the channel code from a codebook. A codebook, carrying all the available channels is maintained and one codeword is transmitted in one  \amj{time/frequency} slot. This scheme, is then assessed using  EC  \amj{model}.   Extensive mathematical and \amj{simulation} evaluation of the proposed scheme shows that, receiver can only decode  codeword when  transmission rate is low.  
\subsubsection{Chase Combining HARQ (CC-HARQ)}
EC  of chase combining HARQ (CC-HARQ) scheme  has been discussed in  \cite{R165}. In this CC-HARQ scheme,  power gain is achieved before decoding a packet. Repeated transmissions are achieved in each packet transmission.  Performance of  CC-HARQ has been compared with  incremental redundancy HARQ (HARQ-IR) based on the  \amj{concept of achievable}  EC. \amj{This analysis shows that,} Type-I HARQ and CC-HARQ are the only two  \amj{retransmission} schemes that can perform  repeated transmission in a single stream.   
\subsubsection{Incremental Redundancy HARQ (HARQ-IR)}
Type-II HARQ is also termed as   \amj{an} incremental redundancy HARQ (HARQ-IR). In this scheme, sender can select either FEC or error detecting parity bits. Work in \cite{R165}, considers  HARQ-IR with the concept of EC. A comparative view of HARQ-IR has also been provided with other versions of HARQ. In HARQ-IR, a complex codeword is first abstracted into simple sub-codewords, and then is transmitted in consecutive time slots, untill the transmission deadline  \amj{expires}.   \amj{In this study, the maximization of EC with}  HARQ-IR has been  \amj{investigated with the required delay-outage probability.}   Authors in \cite{R247,R313}, have provided an-depth analysis of HARQ-IR with delay-sensitive traffic while considering  EC concept.  Impact of  retransmission and other advance schemes of physical layer on  \amj{a}  data link layer especially the queuing performance has also been studied with the EC model. In this work,  physical layer parameters (codewords of HARQ-IR) has been adjusted according to  queuing requirements of  delay-sensitive traffic. \amj{The maximum arrival/source rate has been found with achievable EC to analyze the physical channel with required delay requirements.}  \blue{Statistical QoS provisioning with  HARQ-IR in buffer-aided diamond relay systems has been discussed in \cite{ner3}. Concept of outage effective capacity has been introduced to remove the packets from  buffer that are not successfully received after a certain number of transmission attempts. This EC analysis shows that, the performance of HARQ-IR  is same either using  buffer-aided diamond relaying or decode-and forward relaying.      
}
\subsubsection{Persistent HARQ}
Performance of  persistent HARQ has also been tested while using  EC concept  \cite{R06}.  In this technique,  \amj{a} transmitter persistently resends  packets for a given duration of \amj{a} time (frame). In this scheme, each retransmission scheme is accompanied by \amj{a} reward and is analysed  \amj{with achievable EC with delay requirements.}   It is the only study, that   takes into consideration   \amj{a} detailed analysis of  various versions of HARQ with EC concept.          

\subsubsection{Truncated HARQ}
Authors in \cite{R06}, have  considered     \amj{the mathematical framework based on EC for}  truncated HARQ.  In truncated HARQ, an upper limit is imposed on the number of retransmission schemes. This upper limit is often termed as  ``transmission limit''. With an upper limit on retransmission,  excessive energy for  excessive retransmission has been conserved.   Performance of  \amj{a} truncated HARQ while considering   ``transmission limit''  \amj{with maximum achievable EC has been investigated in detail with different delay-bounds.}      
\subsection{Summary and Insights}
In this section, an in-depth analysis of various retransmission schemes has been discussed while considering the EC model. \amj{ However, to achieve the utlra-reliability and low-latency, retransmission schemes have not been taken into account in some of the mission-critical applications of 5G, such as,  URLLC  \cite{urllc1}.  To support these applications, finite code-length has been used instead of retransmission schemes. This can result into achieving the ultra-reliability and very low latency.  Therefore,  under the stringent delay requirements, in addition to the retransmission schemes, other vistas such as diversity \cite{urllc1}  should be explored to achieve the given reliability.}
    
\section{Open Issues, Challenges, and Future Research Directions}
\label{sec:open-issues}
\amj{As the complexity or size of the networks increases, the mathematical modelling for the EC maximization also tends to become more difficult and complex. For example, the monotonicity and concavity for  the achievable EC in    \blue{multi-carrier} system and  \blue{single-carrier}  system does not necessarily  hold. This could pose a serious challenge for employing the EC to understand the performance of the different networks while providing the statistical QoS requirements.} 

\amj{Extensive work has been done on EC while considering various wireless networks. With the emergence of 5G and beyond, the provision of stringent delay requirements needs to be addressed very carefully. Different applications have different delay requirement (some require very low-delay bound). Using \blue{the} EC to model the delay for various applications and wireless networks with various delay-bounds is a challenging task.  Wireless networks have also their own limitations and constraints such as energy limitation in WSNs. Maximizing  EC while residing within these limitations is  even further a challenging task. After a careful overview of existing work on EC, we have  realized that there are also good potential for the future researchers to use the concept of EC with other state-of-the-art technologies which have also been highlighted below:}

\subsection{Effective Capacity and Ultra-reliable Low Latency Communications (URLLC)}
\amj{International Mobile Telecommunications (IMT)-advance usually concerned about the data rate. However, to accommodate the escalating  demand of emerging future wireless networks, IMT-2020 provides the guidelines for the enhanced mobile broadband (eMBB), URLLC, and massive machine type communications (massive MTC). With latency requirement of $1\rm{ms}$, and $99.999\%$ reliability, URLLC has gained much attention for the mission-critical applications such as Industry 4.0. To meet the stringent delay requirements of URLLC, transmission-delay and queueing-delay should be very small \cite{NOMAmjad}.} 

\amj{Authors in \cite{poor1} have discussed the various tools/metrics such as effective bandwidth, extreme value theorem, and stochastic network calculus  to investigate the techniques in URLLC. In URLLC, the main concern is all about the modelling of the factors with much accuracy that occur very rarely.  Main benefit of using EC to model the delay in URLLC is its simple modelling of the  link performance  \cite{R110}.}  \blue{Ultra latency as well as the reliability have been modeled using the EC model and the short-length codes (short packet communications)  in \cite{ner2}.  Two cases in which the instantaneous  and statistical CSI that are available at  the transmitter are considered to find the bounds on reliability and latency. Using the EC tool to investigation  URLLC with short packet communications still needs further insights from researchers in academia and industry.  Topics such as optimal resource allocation with short packet communications open the door for future research.  }
\subsection{Energy Harvesting Based EC Maximization}
For  wireless networks,  energy is a scarce commodity.  \amj{Conserving}   energy in wireless networks,   has  been a prime goal  for researchers \cite{R97,R126,R128}. With  introduction of energy-harvesting batteries or modules,  performance of wireless networks can be improved \cite{Akhtar15rser}. To the best of authors's knowledge, very limited work has been done on  investigating the effect of  energy-harvesting on the achievable EC of wireless networks. \amj{Depending upon the energy-harvesting source (such as sun in case of solar energy) the arrival of energy-packets is uncertain or not fixed. Therefore, with the introduction of delay requirements this uncertain behavior of arrived energy-packets can  pose a serious challenge to guarantee the low delay communications.    To address this challenge, EC can be used to model the stringent delay requirements and  to investigate the different limitations of energy-harvesting system. This analysis can  also help in changing and improving the design of energy-harvesting modules and transmitters.     } 

\subsection{ \amj{Challenges Related to Effective Capacity Simulation}}
\amj{As compared to the Shannon and ergodic capacity, simulation of the EC is tricky one especially in multi-carrier and multi-user case. As compared to the single user and single-carrier, concavity as well as the monotonicity for EC does not hold   in multiuser and multi-carrier systems \cite{R01}. Therefore, as the complexity of the network increases, the simulations of the EC becomes more and more challenging. Simulations, as well as the mathematical modelling considering the EC, become more complex while considering the complex network scenarios such as the case in heterogeneous networks. }   

\amj{Authors in \cite{R43} have proposed an EC estimation tool named as ``CrEST''. It is the theoretical EC estimation tool for single-hop and single-carrier systems. Different parameters for EC such as delay-bound, queue-length, and others are pre-defined and then the maximization in EC is estimated to assess the performance of the system. However, this tool do not take into consideration the EC maximization in multi-carrier and multi-user systems.}

\subsection{Security and Privacy  Issues in Low Latency Communications} 
Security is  \amj{a} strong pillar in wireless communication for transmitting  common and confidential information. Any breach in the communication security can result into the damage to user data or user privacy.   A well-known technique to provide  security is to use cryptography, however because of security code bits, providing delay-limited communications is even more challenging. EC-based delay analysis in various wireless networks with different security threats has also been studied in  \cite{R120,R136}. \amj{In normal scenarios, EC estimates the delay of information bits in the transmission-buffer. However, in cryptography, in addition to the information bits, there are also the coded bits in the information buffer. In this case, EC takes into consideration the delay of  coded bits only, so the delay analysis of the information bits remains  \blue{unattended}.  This opens the door for the future research to investigate delay-analysis while considering the cryptography techniques.}

\subsection{Effective Capacity of Cognitive Radio Networks (CRNs)} 
\amj{In normal operation of CRNs, SUs have to wait for the PUs to vacate the channel. This can result into some uncertain or  \blue{undefined}  delay depending on the PUs activity patterns. In scenarios where the SUs have also to support the delay-sensitive applications, then the combination of this delay requirement and the undefined delay for accessing the channel can complicate the situation. This challenge needs to be addressed by carefully analysing the delay and performance of the system. For this purpose, EC has been extensively used to understand the delay analysis of the CRNs.}

\subsubsection{Spectrum Sensing in  CRNs and Effective Capacity}  
Among EC-based wireless networks, CRNs have been studied and analyzed extensively with EC metric. Efficient spectrum sensing in CRNs can enhance the performance of   a system.  Transmission of  QoS-aware applications in  CRNs requires efficient sensing and transmission on the part of SUs \cite{R225}.  \amj{Depending on the delay requirements and the PUs activity patterns, in-depth analysis can be performed with the help of EC. This delay-analysis can further be used to design the QoS-aware spectrum sensing approaches to facilitate the delay-sensitive applications.}

\subsubsection{White Space Utilization in  CRNs and Effective Capacity} 
\amj{Depending on  white space utilization, CRNs can further be classified into underlay, overlay, interweave, and hybrid CRNs. Most of the existing work on performance evaluations based on EC metric in CRNs only considers  underlay and overlay CRNs. Depending on the delay requirements, which approach best fits for different delay-sensitive applications should be explored further and this challenge  invites the future researchers to investigate further into this domain. } 

\subsubsection{Spectrum Mobility and TV white space in CRNs and Effective Capacity}
\amj{ In CRNs, simplification of quantification between spectrum mobility, TV white space, spectrum sensing, delay, and PUs activity patterns is a challenging task. Authors in \cite{vasef2012effective}, discuss that, the above mentioned challenge of quantification can be addressed with the help of EC. EC helps in simplifying the performance of the system, which can be used to understand the trade-off between different techniques in CRNs.}

\subsection{ \amj{Challenges Related to Lincensed-Assisted Access in Unlicensed Spectrum}}
\amj{Licensed-assisted access (LAA) in unlicensed band has been witnessed as a promising solution to mitigate the current spectrum crunch for escalating demand of cellular traffic. Provision of QoS in LAA is a challenging task. This challenge is due to the dynamic and heterogeneous nature of LAA and co-existence of the wifi networks. QoS provisioning in LAA can be simplified with the help of EC as has been discussed in \cite{R299}. This EC-based performance analysis invites the future researchers to explore the new vistas related to the unlicensed spectrum sensing, spectrum-mobility, and unlicensed users activity to avoid the interference.} 

\subsection{Optimal Power and Resource Allocation  with Effective Capacity model}
\amj{Resource allocation for  multi-user and multi-carrier systems while considering  \blue{the  EC model} can be challenging.  For single-carrier and  \blue{multi-carrier}, monotonicity and concavity of EC does not necessarily hold. Therefore, with \blue{the  EC model},  resource allocation strategies for  single-carrier systems cannot be simply used for  multi-carrier systems \cite{R01}. Resource allocation with stringent-delay requirements is also a challenging task for the emerging wireless networks. QoS provisioning with optimal resource allocation for the cell-edge users have not been anaylsed in detail \cite{li2018effective}.  Work in  \cite{li2018effective}, is the first attempt to understand the delay requirements for the cell edge users in multi-cell heterogeneous networks. }

\subsection{\amj{Effective Capacity of Heterogeneous Networks}}
\amj{Heterogeneous networks consist of very diverse network components/architecture. These networks have also had to support the diverse delay-sensitive applications with different delay requirements. Therefore, provisioning and analysing the delay for the heterogeneous networks is a challenging task. EC is a useful tool that can simplify the performance analysis of the complex and diverse networks such as heterogeneous networks. EC is a flexible tool that can also provide the clean slate to understand the different  delay requirements (with different delay-bound) for the diverse range of networks \cite{R110}.}

%
%
%

\section{Conclusion}
\label{sec:conclusion}
Quality of service (QoS)-aware and delay-sensitive real-time applications require wireless channel models that can incorporate  QoS-aware evaluation metrics such as delay, data rate, and delay-violation probability. Existing physical-layer channel models  do not consider  QoS metrics. To address this issue, QoS-aware link-layer  wireless channel model named as  ``effective capacity (EC)'' has been proposed. In this paper, we have provided a comprehensive survey of \blue{the  EC model}  with its \amj{state-of-the-art} work in different wireless networks.  How EC metric can be used for testing the  \amj{performance} of various wireless networks, has been surveyed in detail in this paper.   Five different case studies involving   EC concept in cellular networks, device-to-device (D2D) communications, full-duplex (FD) communications, peer-to-peer streaming, and visible light communications have been presented.  Various QoS-aware and delay-sensitive applications such as voice, video, and medical applications analysed by  \blue{the  EC model} have also been surveyed. EC-based delay  \amj{analysis}  in wireless networks under different fading models such as stochastic, generalized, dimension-based, supplementary, and futuristic fading models  have  been provided.  Among these fading channels Rayleigh fading model has been extensively evaluated with EC metric. Concept of EC  in different networks such as CRNs, relay networks, cellular networks, mesh networks, and adhoc networks  \amj{have}    been explored in  detail. This paper also covers the achievable EC in FD communications and various retransmission schemes. In the last, we have concluded this paper by outlining some open issues and future research directions related to \blue{the  EC model} in various existing and advance future wireless networks.

%
%


\end{document}